\documentclass[numberedappendix]{emulateapj}
\newcommand{\Mb}{M_\mathrm{B}}
\newcommand{\Mr}{M_\mathrm{r}}
\newcommand{\MV}{M_\mathrm{V}}
\newcommand{\mb}{m_\mathrm{B}}
\newcommand{\mr}{m_\mathrm{r}}

\newcommand{\smap}{a_\mathrm{p}}

\newcommand{\ahlr}{a_\mathrm{hl,r}}
\newcommand\expall{\emph{exp}}
\newcommand\expA{\emph{exp-1}}
\newcommand\expB{\emph{exp-2}}
\newcommand\burst{\emph{burst}}

\newcommand\cbten{\emph{const-b100}}

\newcommand\uriz{$u-r$ vs.\ $i-z$}
\newcommand\griz{$g-r$ vs.\ $i-z$}

\slugcomment{Accepted for publication by The Astronomical Journal}
\shorttitle{Virgo early-type dwarfs. IV. Color-magnitude}
\shortauthors{Lisker, Grebel, \& Binggeli}
\begin{document}
 
\title{Virgo cluster early-type dwarf galaxies with the Sloan Digital
  Sky Survey.\\
IV. The color-magnitude relation}

\author{
Thorsten Lisker\altaffilmark{1,2},
Eva K. Grebel\altaffilmark{1,2}, 
and Bruno Binggeli\altaffilmark{2}
}
\altaffiltext{1}{
  Astronomisches Rechen-Institut, Centre for Astronomy of Heidelberg
  University (ZAH), Moenchhofstrasse 12-14, D-69120 Heidelberg, Germany
}
\altaffiltext{2}{
Astronomical Institute, Dept.\ of Physics and Astronomy,
  University of Basel, Venusstrasse 7, CH-4102 Binningen, Switzerland\
}
\email{TL@x-astro.net}

\begin{abstract}
We present an analysis of the optical colors of 413 Virgo cluster
early-type dwarf galaxies (dEs), based on Sloan Digital Sky Survey
imaging data. Our study comprises (1) a comparison of the
color-magnitude relation (CMR) of the different dE subclasses that
we identified in Paper III of this series, (2) a comparison of the shape
of the CMR in low and high-density regions, (3) an analysis of the
scatter of the CMR, and (4) an interpretation of the observed colors
with ages and metallicities from population synthesis models. We find
that the CMRs of nucleated (\,dE(N)\,) and non-nucleated dEs (\,dE(nN)\,) 
are significantly different from each other, with
similar colors at fainter magnitudes ($\mr\gtrsim 17$ mag), but
increasingly redder colors of the dE(N)s at brighter
magnitudes. We interpret this with older ages and/or higher metallicities of the brighter dE(N)s. The dEs with disk features have
similar colors as the dE(N)s and seem to be only slightly younger and/or
less metal-rich on
average.
 Furthermore, we find a small but significant dependence of
the CMR on local projected galaxy number density, consistently
seen in all of $u-r$, $g-r$, and $g-i$, and weakly $i-z$.
We deduce that a significant intrinsic color scatter of the CMR is
present, even when allowing for a distance spread
of our galaxies. No increase of the CMR scatter at fainter magnitudes is
observed down to $\mr \approx 17$ mag ($\Mr \approx -14$ mag). The color
residuals, i.e., the offsets of the data points from the linear fit to the
CMR, are clearly correlated with each other in all colors for the dE(N)s
and for the full dE sample, implying that, at a given magnitude, a galaxy
with an older stellar population than average typically also exhibits a
higher metallicity than average.
Given the observational data for Virgo dEs presented here and in the
previous papers of this series, we conclude that there must be at least
two different formation channels for early-type dwarfs in order to explain the
heterogeneity of this class of galaxy.
\end{abstract}
 
\keywords{
 galaxies: dwarf ---
 galaxies: evolution ---
 galaxies: stellar content ---
 galaxies: fundamental parameters ---
 galaxies: photometry ---
 galaxies: clusters: individual (Virgo)
}
 
%________________________________________________________________

\section{Introduction }
\label{sec:int}

%I
It has long been known that a close correlation exists between
the colors and luminosities of early-type galaxies, in the sense that more
luminous objects have redder colors
\citep[e.g.,][]{bau59,deV61,fab73,sanvis78a,cal83}.
%II
A striking observation is the universality of the color-magnitude relation
(CMR): it was found to
be equal, within the measurement errors, for E and S0 galaxies within
clusters, groups, or the field \citep{fab73,sanvis78a,sanvis78b,bow92},
leading \citet{fab73} 
to state that the colors of elliptical galaxies ``are independent of
all physical properties studied other than
luminosity''. \citeauthor{fab73} also showed that a similar
relation exists between the strength of spectral absorption features
and luminosity, which basically is the spectroscopic analogue to the
CMR. She interpreted the CMR as a trend of increasing
metallicity with luminosity, which today is still considered to be the
primary determinant of the CMR \citep[e.g.,][]{kod97,cha06}. This can
be understood with a higher binding energy per unit mass of gas in more
massive galaxies, leading to stronger enrichment of the stellar
populations. 
Recently, \citet{ber03IV} showed that color seems to correlate even
 more strongly with velocity dispersion than it does with
 luminosity. This implies that the CMR itself is most likely
just a combination of the relation of luminosity and velocity
dispersion \citep[``Faber-Jackson relation'',][]{fab76} and that of color and velocity
 dispersion. As \citet{mat05} point out, the latter might hint at
 a more fundamental relation, namely between galaxy metallicity and
 mass.

%III
 While this trend includes
the dEs, there has been disagreement about whether or not they follow
the same CMR as the giant ellipticals. \citet{deV61} found the
dwarfs to be ``systematically bluer'' than the giants (but nevertheless
following a CMR), whereas \citet{cal83} reported a linear CMR over a range of
$-15 \le \MV \le -23$ mag. However, his Figure~3 actually suggests that
the slope of the CMR might indeed be slightly different for the dEs, in
the same way as the results of \citet{deV61} suggested (i.e., with
decreasing magnitude, dwarfs become bluer more rapidly than giants do).
%IV
 \cite{mat05} found a change in the slope of the
 Faber-Jackson relation for ``faint early-type'' galaxies --- here,
 ``faint'' means $-17.3 \le \Mb \le -20.5$ mag, thus reaching only
 slightly into the dwarf regime. \citet{der05} found an
 even larger difference in slope for a sample of 15 dEs, but argued that 
 this is consistent with theoretical models, due to the dynamical
 response to starburst-induced mass loss, which is stronger for objects
 of lower mass. Significant constraints to such models could only be provided
with a better understanding of dE formation. In contrast to giant
 ellipticals, formation mechanisms proposed for dEs in clusters are
 typically not based on an early formation epoch, but rather, on infall and
subsequent transformation of late-type galaxies through gas-stripping and
 tidally enhanced star formation
\citep[e.g.,][]{dav88,moo96,vZe04a,sab05,del07}.

%V
After having established a subdivision scheme of Virgo cluster early-type dwarf (dE)
galaxies into subclasses with different shapes and distributions
\citep[Paper III of this series]{p3},
we can now proceed to the next logical step,
namely to exploiting the wealth of data provided by the Sloan
Digital Sky Survey (SDSS) Data Release 5 \citep[DR5,][]{sdssdr5} by a
multicolor analysis of our sample of 413 dEs. However, the
colors of different dEs, or of dEs of different subclasses, can not
straightforwardly be compared with 
each other: the existence of the CMR
requires such a comparison to be done either at
fixed magnitude or with a correction for magnitude differences. For
this reason, we shall explore the  dE colors
mainly through an analysis of their CMR.

%VI
One would naively expect that, if
two given dE subclasses formed through different mechanisms, their
resulting CMRs should display differences as well, since the relation
of galaxy mass to the properties of its
stellar population should depend to some
extent on how and when the latter was formed. On the other hand, the
apparent universality of the CMR -- mainly defined for giant
ellipticals --
would seem to argue against such differences.
%VII
%VIII
\citet{conIII} found 
considerable scatter of the CMR of dEs in the Perseus cluster at
magnitudes $\Mb \ge -15$ mag, apparently caused 
by two different sequences of dwarfs in color-magnitude space.
 They argue that the
early-type dwarfs must have multiple origins --- similar to the conclusion
of \citet{pog01} about an apparent bimodal metallicity distribution of
Coma cluster dwarfs that might hint at more than one formation channel.
We should be able to test the presence of such a bimodality in more detail, given our
``preparatory work'', namely the separation of dE subclasses that have
different shapes and distributions (Paper III). Moreover,
the SDSS multicolor data enable us to construct CMRs in more than one
color, and also to analyze our galaxies in color-color space,
thereby translating colors into ages and metallicities. Since 
\citet{rak04} found the dE(nN)s in the Coma and Fornax clusters to
be younger and to have a higher metallicity than the dE(N)s, we can
perform a similar analysis for our Virgo cluster galaxies,
allowing us to test the similarity of dE populations
of different clusters.
%

%%%%%%%%%%%%%%%%%%%%%%%%%%%%%%%%%%
\section{Data}
\label{sec:data}

  The SDSS DR5 covers
  all galaxies listed in the Virgo cluster catalog \citep[VCC,][]{vcc},
  except for an
  approximately $2\arcdeg \times 2\fdg5$ area at $\alpha\approx 186\fdg2$,
  $\delta\approx +5\fdg0$. It provides reduced images taken in the 
  $u$, $g$, $r$, $i$, and $z$ bands with an effective exposure time of
  $54 \rm{s}$ in each band \citep[see also][]{sdssedr}, as well as the
  necessary parameters to flux-calibrate them. The pixel scale
  of $0\farcs396$ 
  corresponds to a physical size of $30$ pc at our adopted Virgo
  cluster distance of $d=15.85$ Mpc \citep[distance modulus
  $m-M=31.0$ mag;][]{vdB96,gra99}, which we use throughout. 

  The SDSS imaging camera
  \citep{sdsscamera} takes data in drift-scanning 
  mode nearly simultaneously in the five photometric bands, and thus
  combines very homogeneous multicolor photometry 
  with large area coverage and sufficient depth to
  enable a systematic analysis of dEs.
  The images have an absolute astrometric accuracy of rms$\le 0\farcs1$ per
  coordinate, and a relative accuracy between the $r$ band 
  and each of the other bands of less than $0.1$ pixels
  \citep{sdssastrometry}. They can thus easily be aligned using
  their astrometric calibration and need not be registered
  manually. Furthermore, adjacent SDSS images can be accurately combined,
  allowing the extraction of subimages that fully cover a given object,
  even if the latter lies at the edge of an SDSS image. 

  The rms of the noise per pixel corresponds to a
  surface brightness of approximately $24.2$ mag arcsec$^{-2}$ in the
  $u$ band, $24.7$ in $g$, $24.4$ in $r$, $23.9$ in $i$, and $22.4$ in $z$.
  The typical total signal-to-noise ratio 
  (S/N) of a bright dE ($\mb \approx 14$ mag) amounts to about $1000$ 
  in $r$, within an aperture radius of approximately two half-light
  radii. For a faint dE ($\mb \approx 18$ mag) this 
  value is typically about $50$. While the S/N in $g$ and $i$ is
  similar, it is several times lower in $z$ and almost ten times lower in
  $u$.

  Since the sky level on the SDSS images can vary by some tenths of the noise
  level across an image, it is not sufficient to subtract only a single sky
  flux value from each image. We therefore determined the sky flux
  distribution across a given image using a thorough procedure, as described
  in detail in Paper III. The sky-subtracted images were then
  flux-calibrated and corrected for Galactic extinction \citep{sch98}. We also
  correct for the reported\footnotemark[3]
\footnotetext[3]{~See http://www.sdss.org/dr5/algorithms/fluxcal.html}
  SDSS zeropoint offsets in
  $u$ and $z$ from the AB system \citep{ABsystem}:
\begin{equation}
u_{\rm AB} = u_{\rm SDSS} - 0.04\ \rm{mag}
\end{equation}
\begin{equation}
z_{\rm AB} = z_{\rm SDSS} + 0.02\ \rm{mag}
\end{equation}

%%%%%%%%%%%%%%%%%%%%%%%%%%%%%%%%%%
\section{Sample and basic measurements}
\label{sec:sample}

Our working sample of Virgo cluster dEs contains 413 certain cluster members
 that were initially classified as early-type dwarfs in the VCC (``dE'' or
 ``dS0'', including uncertain candidates), that are
 brighter than $\mb\le18.0$ mag, that passed our visual examination for
 confusion with possible dwarf irregulars, and for which a Petrosian radius
 \citep{pet76} could be derived. The details of our sample selection are
 described in 
 Paper III.

 For each galaxy, we determined a ``Petrosian semimajor axis'' (hereafter
 Petrosian SMA, $\smap$), i.e.,
 we use ellipses instead of circles in the calculation of the Petrosian radius
 \citep[see, e.g.,][]{lot04}. 
 The total flux in the $r$ band was measured within $a = 2\,\smap$, yielding a
 value for the half-light semimajor axis,  $\ahlr$. Axial ratio and
 position angle were then determined through an isophotal fit at
  $a = 2\,\ahlr$. The details of this process are outlined in
 Paper III. Disturbing foreground or background objects were
 properly masked, and the area blocked by these masks was taken into account
 in the determination of the total $r$-band flux. We also decided to mask the
 nuclei, if 
 present, in order to guarantee that a nucleus with a different color
 than its host galaxy would not affect our measurements of the
 inner galaxy colors. For each galaxy and band, we measured the flux within
 three elliptical apertures:
$a \le 0.5\,\ahlr$ (``small aperture''), $a \le \ahlr$ (``intermediate
aperture'' or half-light aperture), and $a \le 2\,\ahlr$ (``large aperture''). 

%%%%%%%%%%%%%%%%%%%%%%%%%%%%%%%%%%
\section{Error estimation}
\label{sec:errors}

The actual errors on the measured fluxes are a combination from several
different sources of uncertainties, which we attempt to estimate as
realistically as possible.
The noise level in every image is measured as the standard deviation around
the mean of all unmasked pixels, clipped five times iteratively at $3
\sigma$. For each galaxy, band, and aperture, we then adopt the
following uncertainties of a given flux measurement:
\begin{itemize}
\item The ``S/N uncertainty'' is the inverse S/N value.
\item The ``sky level uncertainty'' is derived from the uncertainty in the 
sky level determination per pixel, which is taken to be 0.2\% of the noise
level: the mean flux value of all unmasked
pixels after sky subtraction typically deviates from zero by this value or
less.
\item The uncertainty in the determination of the Petrosian SMA (``Petrosian
  uncertainty'') is estimated to be of the same order as the sky level
  uncertainty, which we therefore simply count
  twice in our calculation of total errors. We elaborate on this uncertainty
  in more detail in Appendix~\ref{app:errors}.
\item The uncertainty in the photometric calibration (``calibration
  uncertainty'') is reported\footnotemark[3] as $0.02$ mag in $g$, $r$, and
  $i$, and $0.03$ mag in $u$ and $z$.
\item The natural red leak of the $u$ filter causes the ``$u$-leak
  uncertainty'' of $0.02$ mag\footnotemark[3].
\end{itemize}
For each flux measurement $f_{\rm x}$ in a given band $x$, we combine
all the different uncertainties to a single uncertainty $\Delta f_{\rm
  x}$. This is done by adding the individual uncertainties quadratically, as
exemplified below for the $u$ band.
%\begin{equation}
\begin{eqnarray}
\Delta f_{\rm u}
= 
(
(\frac{\sigma_{\rm u}\cdot \sqrt{N_{\rm
            pix}}}{f_{\rm u}})^2
+
2\cdot (\frac{0.002\,\sigma_{\rm u}\cdot N_{\rm pix}}{f_{\rm u}})^2
+\nonumber\\
(10^{0.4\cdot 0.03}-1)^2
+
(10^{0.4\cdot 0.02}-1)^2
)^{0.5}
\end{eqnarray}
%\end{equation}
Here, $\sigma_{\rm u}$ denotes the noise level per
pixel, as described above. $N_{\rm pix}$ is the number of pixels
included in the given aperture. The first term is the S/N uncertainty, the
second term is the combination of sky level uncertainty and Petrosian
uncertainty, and the last two terms are the calibration
uncertainty and the $u$-leak uncertainty, respectively.

Errors on measured color values are calculated by
adding the relative flux errors from each band quadratically, and
converting them to magnitudes. An example is given below for the $u-g$ color
error, $\Delta(u-g)$,
calculated from the $u$ band flux $f_{\rm u}$, the $g$ band flux $f_{\rm
  g}$, and their respective errors.
\begin{equation}
\label{eq:errmagshort}
\Delta(u-g) = -2.5\log\left(1-\sqrt{ \left(\frac{\Delta f_{\rm u}}{f_{\rm
          u}}\right)^2 + \left(\frac{\Delta f_{\rm g}}{f_{\rm g}}\right)^2
}\,\right) 
\end{equation}
Equation \ref{eq:errmagshort} would actually yield only the errors on one side,
whereas the errors on the other side would be calculated from
$-2.5\log(1+\sqrt{...})$, and would consequently be smaller. However, we
  prefer the conservative approach to use the larger errors for both
  sides. Galaxies for which the relative flux error becomes $\ge$1
  are excluded from the respective diagrams or calculations, since the
  argument to the logarithm in Equation \ref{eq:errmagshort} would become $\le$0,
  denoting an uncertainty that is too large to be useful. This occurs only for
  five of the fainter objects, for colors that include the $u$ band.

%%%%%%%%%%%%%%%%%%%%%%%%%%%%%%%%%%
\section{Statistical techniques}
\label{sec:statistic}

\subsection{Linear fitting of color-magnitude relations}
\label{sec:sub_fitting}

Since color-magnitude relations of early-type galaxies are found to be
linear or nearly linear \citep[e.g.,][]{bow92,ber03IV}
we describe our relations by fitting a straight line to the data points.
A common approach is  least-squares fitting, which, however, can sometimes lead
to undesired results, e.g., a ``best fit'' with a slope that differs
significantly from that defined by the bulk of data points \citep[see their
  Figure~15.7.1]{nr}.
We therefore prefer to apply a robust fitting
technique using a so-called M-estimate \citep{nr}, namely the mean
\emph{absolute} deviation (rather than the mean square deviation used for 
 least-squares fitting). See, e.g., \citet{cha06} for an earlier application
 of this technique to color-magnitude relations.

 We thus have to minimize the function
\begin{equation}
\mu(a,b) = \displaystyle\sum_{i=0}^{N-1}\left|\frac{c_{\rm
    i}-a-b\cdot m_{\rm i}}{\sigma_{\rm i}}\right|
\label{eq:m_est}
\end{equation}
where $a$ and $b$ are the zeropoint and the slope of the relation,
respectively, $m_{\rm i}$ and $c_{\rm i}$ are magnitude and color of the i-th
galaxy, and $\sigma_{\rm i}$ is the corresponding
measurement uncertainty, or alternatively, any kind of inverse weight.
We set $\sigma_{\rm i}$ equal to the respective color error, neglecting
magnitude errors, since they are small compared to the range of magnitudes
   that is considered (compare, e.g., Figure~\ref{fig:cmr}), and since they
   are also closely correlated with the color errors. For some cases we
   tested $\sigma_{\rm i}=\sqrt{(\Delta m_{\rm
   i})^2+(\Delta c_{\rm i})^2}$, i.e., quadratically adding
   magnitude and color error. We found no significant differences in the
   resulting best fit.
The fitting process is repeated two times iteratively, excluding data points
 that are offset from the fitted line by more than three times the derived rms
 scatter. 

%%%%%%%%%%%%%%%%%%%%%%%%%%%%%%%%%%
\subsection{Statistical comparisons of color-magnitude relations}
\label{sec:sub_statcompare}

In the following sections, we aim at performing
\emph{quantitative}, statistical comparisons of two given color-magnitude
relations. Here, we briefly outline our method for such a comparison, and
present a more detailed description in Appendix \ref{app:statistics}. 

We decided on an approach that takes into account both the data points themselves and
the linear fits to them. For two given sets of data points A and B, we compute
the color residuals of both sets about the CMR A and statistically compare
these residuals through a \mbox{K-S} test and a Student's t-test for unequal
variances. This is done analogously for the residuals about the CMR B. The
final probability for a common underlying distribution from the two \mbox{K-S} tests
is then derived by averaging both values of the \mbox{K-S} statistic $D$
\citep[see][]{nr}. The final probability from the two t-tests is adopted
to be the larger of the two individual probabilities.

In the course of the
paper, we will usually simply refer to the higher of the probabilities
from the \mbox{K-S} test and the t-test. We typically consider two CMRs to be
``significantly'' different if the statistical comparison yields probabilities
below a few percent.

%%%%%%%%%%%%%%%%%%%%%%%%%%%%%%%%%%
\section{Color-magnitude relations of early-type dwarf subclasses}
\label{sec:subclasses}
 
In Paper III we established a subdivision scheme for the
dEs that comprises several subclasses with different properties: dEs
with disk features like spiral arms or bars (dE(di)s, \citealt{p1}, Paper I), dEs with
blue centers (dE(bc)s, \citealt{p2}, Paper II) caused by recent or ongoing central star
formation, bright and faint ordinary dEs (i.e., not displaying disk
substructure or a blue center) that are nucleated (dE(N)s),
and bright and faint ordinary dEs that have no nucleus or only a weak
nucleus that is below the detection limit of the VCC (dE(nN)s). The
bright dE(nN)s, dE(di)s, and dE(bc)s are shaped like thick
disks and show no central clustering, while the faint dE(nN)s and the
dE(N)s have rounder shapes, and the dE(N)s are more strongly
concentrated towards the cluster center. These results define a
morphology-density relation \emph{within} the dE class, and we now seek
to investigate whether this is also correlated with color.

\begin{figure}
  \epsscale{1.0}
 \plotone{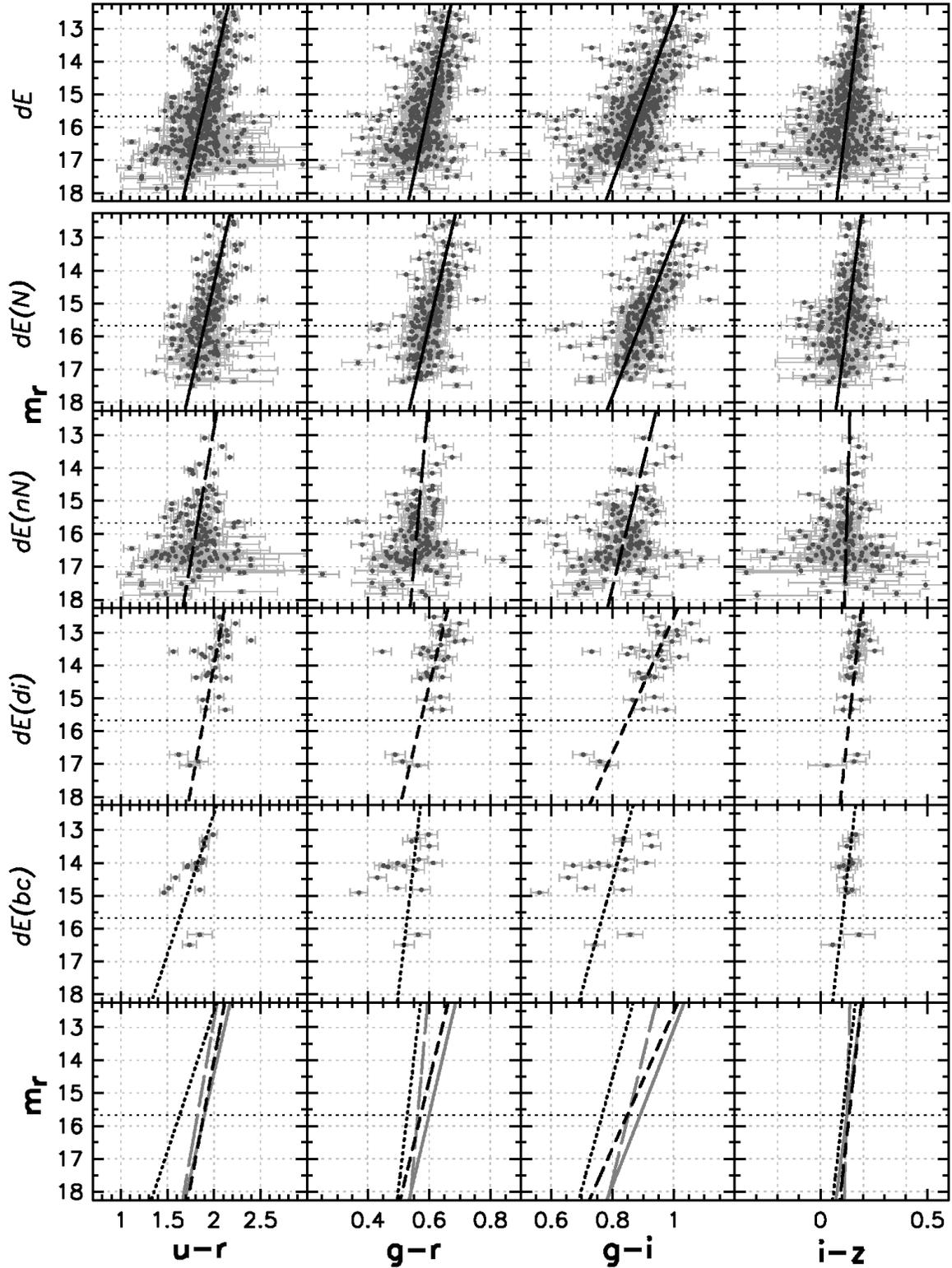} %{cmr.eps}
  \caption{ Color-magnitude relations of dE subclasses.
    Shown are the relations of $r$ magnitude with $u-r$, $g-r$, $g-i$, and
    $i-z$ color for the different dE subclasses (different rows, as
    labelled on the left-hand side) and the full dE sample excluding
    dE(bc)s (top row), using the half-light aperture. The
    linear fits to the CMR of each subclass are shown as lines of different style,
    and are plotted again in the bottom row for comparison. There, two
    lines are shown in grey, to allow a better visual distinction. Both
    color and magnitude errors are shown; however, the latter are in most
    cases smaller than the size of the data points.
  }
  \label{fig:cmr}
\end{figure}

In Figure~\ref{fig:cmr} we present the relations of
color and $r$ magnitude separately for dE(N)s, dE(nN)s, dE(di)s, and
dE(bc)s, as well as for the full dE sample excluding dE(bc)s.
Our choice of presented colors
relies partly on the considerations outlined in Section~\ref{sec:sub_modelsetup}:
$u-r$ is mainly age-sensitive, $i-z$ is metallicity-sensitive, $g-i$ provides
the largest wavelength baseline within the three high-S/N SDSS bands ($g$,
$r$, $i$), and $g-r$ is an alternative to $u-r$, since the smaller
wavelength range of $g-r$ is counterbalanced by the significantly smaller
errors in $g$ as compared to $u$.
The horizontal dotted 
lines denote the separation between our bright and faint subsamples,
which were divided at the median $r$ brightness of our full sample, $\mr=15.67$
mag (Paper III; corresponding to $\Mr=-15.33$ when adopting $m-M=31.0$).
The CMRs of the dE subclasses are compared to each other in the lowermost row
of the figure.

The resulting parameters of the
corresponding linear fits are given in Table~\ref{tab:cmr}. While the colors
presented in Figure~\ref{fig:cmr} were measured within the half-light
aperture (as defined in Section~\ref{sec:sample}), Table~\ref{tab:cmr} also
includes the CMR-parameters for the small 
and large apertures. These will be taken into account in
Section~\ref{sec:sub_modelapplication}, where
systematic radial stellar population gradients will be discussed.

First, we focus on the dE(bc)s. From Paper II,
we 
expect that the colors beyond $a > 0.5\,\ahlr$ are on average only slightly
affected by the positive central color gradient, and that the colors are not
noticeably affected beyond  $a > \ahlr$. However, in all of $u-r$, $g-r$, and
$g-i$, the dE(bc)s are somewhat bluer than the other dE subclasses. Note that
no precise statements can be made about the slope of their CMR, since it is
constrained at the faint end by only two data points. Nevertheless, it can be
seen that the scatter in $g-r$ and $g-i$ is somewhat larger than that of the
other subclasses at the same magnitudes. This is confirmed by the respective
values of the rms scatter, which are larger by a factor of $\sim$1.5 for the dE(bc)s. In
contrast, the dE(bc)s follow a surprisingly tight relation in $i-z$: the
scatter is a factor of two lower than the rms of the measurement
errors\footnote[4]{~This, along with the tight relation of the dE(di)s,
  confirms that we did not underestimate our 
  photometric errors} (see Table~\ref{tab:cmr}), and likewise, it is lower than
the scatter of all other subclasses in this bin. At larger radii, the relation
seems to turn around, i.e., such that brighter galaxies are bluer, but again,
the small sample size does not allow robust conclusions. Nevertheless, a trend
for the CMR to turn around is also seen for the dE(N)s and the dE(nN)s.

The CMR of the dE(di)s is similar to that of the dE(N)s in all
of $u-r$, $g-r$, and $g-i$, although it is always slightly bluer. According to
our statistical tests, as described in 
Section~\ref{sec:sub_statcompare} and presented in Table~\ref{tab:tut}, this
small 
difference is nevertheless significant in $g-i$: for all apertures, the
probability of a common underlying distribution function\footnote[5]{~
While we always consider the probabilities from both the \mbox{K-S} test and
the t-test, we recall that the t-test compares the \emph{means} of the
residuals about the CMR, not the distributions themselves. For simplicity,
though, we shall continue speaking of the ``probability of a common
underlying distribution''.} with the
dE(N)s is $\le$1.6\%. For the small and intermediate apertures of $g-r$, it
is $\le$2.3\%, and 
for the small aperture of $u-r$ it is $\le$1.7\%. We note, though, that
these values should be taken with a grain of salt,
since the linear fit to the dE(di)s is constrained by only three data
points at the faint end, and since the sample is of only moderate
size. Nevertheless, the slight blueward
offset of the dE(di)s from the dE(N)s is consistently seen in all the above
bands.
In $i-z$, however, the dE(di)s basically follow the same relation
as the dE(N)s. For the large aperture, the dE(di)s are slightly
redder, but the difference is not significant (see Table~\ref{tab:tut}).

\begin{figure}
  \epsscale{1.0}
 \plotone{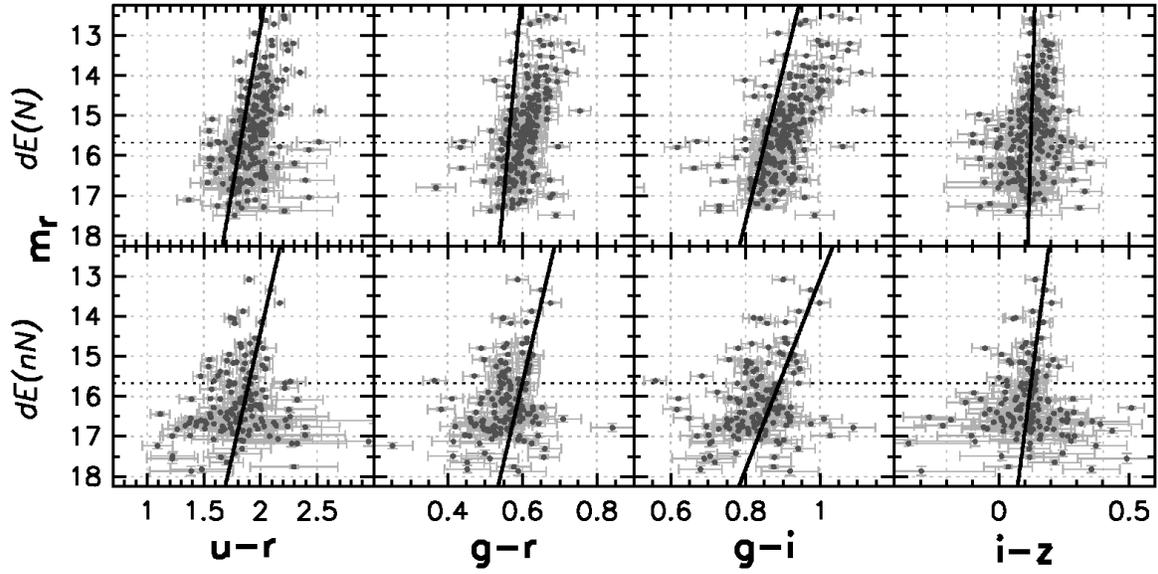} %{cmr3.eps}
  \caption{ Comparison of the CMR of dE(N)s and dE(nN)s.
    Similar to the second and third row of Figure~\ref{fig:cmr}, but now
    overplotting the fit to the 
    dE(nN)s over the data points of the dE(N)s, and vice versa.
  }
  \label{fig:cmr3}
\end{figure}

The dE(nN)s follow a steeper CMR than the dE(N)s or the dE(di)s in all
colors and apertures, including $i-z$ even though the difference is
small there. They lie at about the same color values at the faint end of the
sample, but become red less rapidly with increasing magnitude. The difference between
the CMRs of the dE(nN)s and the dE(N)s is significant in all apertures of
$g-r$ (probability of common distribution $\le$0.5\%) and $g-i$ ($0.0\%$). It
is, however, not significant in $i-z$ ($\ge$15\%), and hardly significant in
$u-r$ ($\le$9\%). To illustrate these differences in a more
obvious way than Figure~\ref{fig:cmr} does, we overplot in
Figure~\ref{fig:cmr3} the linear fit of the respective \emph{other} subclass
over 
the data points of dE(N)s and dE(nN)s --- the discrepancy in $g-r$ and
$g-i$ is clearly seen. 

In order to further quantify the above considerations, one would want to
compare average color values of the subsamples. The problem with this
approach is that
the samples are not distributed equally in magnitude; the average colors
could thus come out different just because of the existence of a
color-magnitude relation. Therefore, we 
compare instead the color values of the linear fits to the CMRs at a fixed
magnitude. We choose two magnitude values for this comparison,
namely the median $r$ magnitude of our bright dEs, ${m_\mathrm{r,bri.}}=14.77$
mag (${M_\mathrm{r,bri.}}=-16.23$)
and the median $r$ magnitude of our faint dEs,
${m_\mathrm{r,fai.}}=16.51$ mag (${M_\mathrm{r,fai.}}=-14.49$), which we
refer to hereafter as the bright and faint ``reference magnitude'',
respectively. The corresponding color values for the 
different subclasses are given in Table~\ref{tab:medbrifai}.

The values so derived reflect our above comparisons of the various CMRs: the dE(N)s are
redder than the dE(nN)s in $u-r$, $g-r$, and $g-i$ at both the bright and the
faint reference magnitudes, with the difference being smaller for the
latter. In $i-z$, the dE(N)s are still redder at the bright reference
magnitude, whereas no clear difference is present at the faint reference
magnitude, also given the rather large scatter of the CMR.  The dE(di)s,
which we consider only at the bright reference magnitude, are intermediate
between the dE(nN)s and the dE(N)s in $u-r$, $g-r$, and $g-i$. In $i-z$,
their colors are very similar to  those of the dE(N)s, again reflecting
our above analysis of the CMRs. In Section~\ref{sec:stelpop}, we will
attempt to interpret these color values in terms of age and/or metallicity
differences.

\begin{figure}
  \epsscale{1.0}
 \plotone{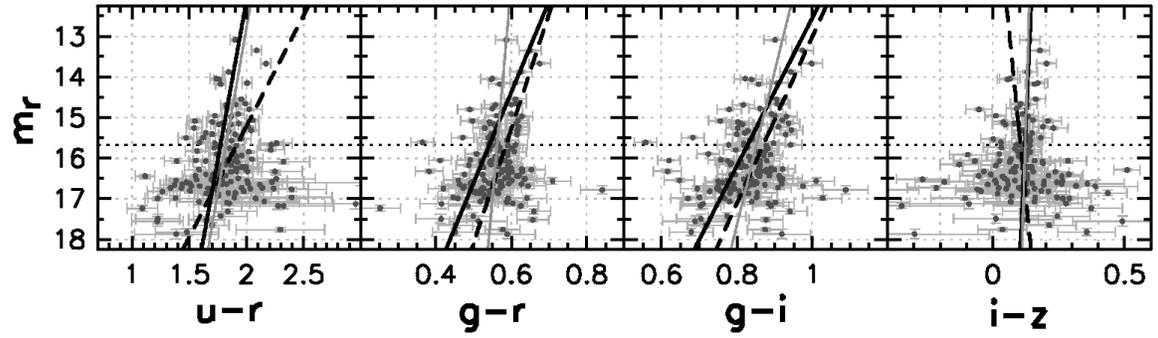} %{cmrdEnN.eps}
  \caption{ Comparison of the CMR of bright and faint dE(nN)s. Similar to
  the third row of
  Figure~\ref{fig:cmr}, but now showing the fit to the full dE(nN) sample
  as grey line, the fit to only the bright subsample as black solid line,
  and the fit to only the faint subsample as black dashed  line.
  }
  \label{fig:cmrdEnN}
\end{figure}

We need to point out one possible caveat with the above comparisons: in
Paper III we found significant differences in both the shapes and the
distributions of faint and bright dE(nN)s. It might thus be not
appropriate to fit a single line through their combined sample. Therefore,
we show in Figure~\ref{fig:cmrdEnN} separate fits to the  CMRs of the
bright and faint dEnN subsamples. While these do differ from the fit to
the full sample, they are rather poorly constrained: the range in
magnitude covered by each subsample is not very large compared with the
scatter of the data points. The linear fits might thus be not reliable
enough to allow a robust statistical comparison. Nevertheless, a
comparison with a similar fit to the bright subsample of the dE(N)s shows
that the bright dE(nN)s are still significantly bluer than the (bright)
dE(N)s in $u-r$, $g-r$, and $g-i$ at the bright reference magnitude
(see the two rightmost columns of Table~\ref{tab:medbrifai}).

\section{Color-magnitude relations of density-selected samples}
\label{sec:density}

A weak dependence of the stellar population ages of 
early-type galaxies on local environmental density has recently been
reported by \citet{ber06}, based on a study of Lick indices. These
authors found early-type galaxies in high-density environments to be
slightly older than at lower densities. Our large
sample size allows us to test whether this is the case for dEs as well ---
we can of course only probe a cluster environment, but the range in local
densities (see, e.g., Paper III) should be large enough to find
such an effect, if present.

We thus subdivide the sample of dE(N)s, dE(nN)s, and our full dE sample
 (excluding dE(bc)s)
 at their respective median local projected
densities. These are, in units of the logarithm of the number of galaxies per
square degree, 1.374 for the dE(N)s, 1.198 for the dE(nN)s, and 1.293 for the
 full dE sample. It can be seen from these values that the dE(N)s populate
 on average regions of higher density   than the other subclasses,
 reflecting their centrally clustered spatial distribution, which is not
 seen for the other subclasses (Paper III). 
 As described in the previous papers of this series,
 local projected density is calculated by defining a circular area around 
each galaxy that includes its ten nearest neighbour cluster members,
 independent of galaxy type \citep{dre80,bin87}.

\begin{figure}
  \epsscale{1.0}
 \plotone{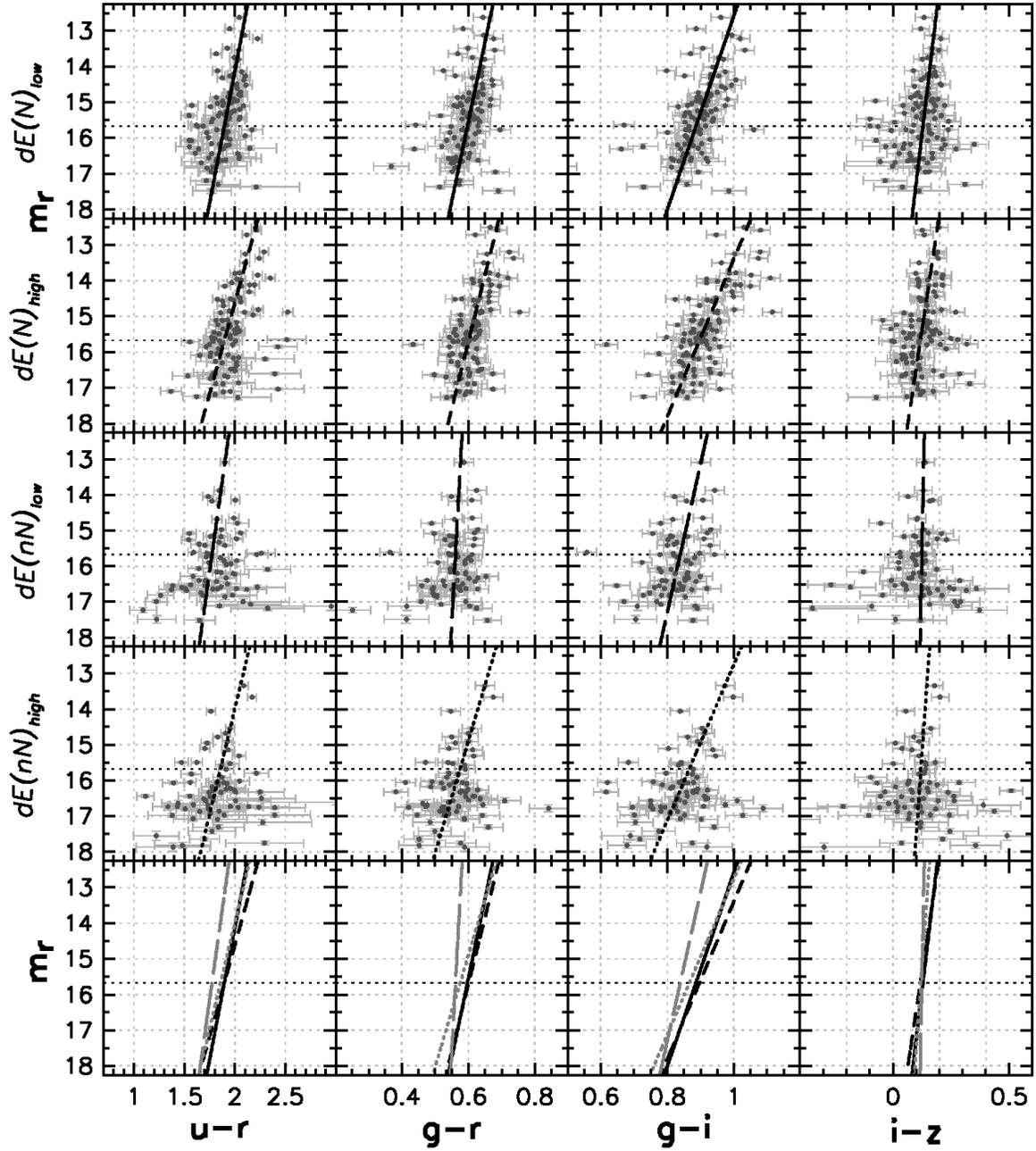} %{cmrdens.eps}
  \caption{ Color-magnitude relations for different densities.
    Similar to Figure~\ref{fig:cmr}, but showing the CMR of the low-density
  and high-density subsamples of the dE(N)s and the dE(nN)s, as labelled on
  the left-hand side. Again, two of the lines are shown in grey in the
    bottom row, to allow a better visual distinction.
  }
  \label{fig:cmrdens}
\end{figure}

\begin{figure}
  \epsscale{1.0}
 \plotone{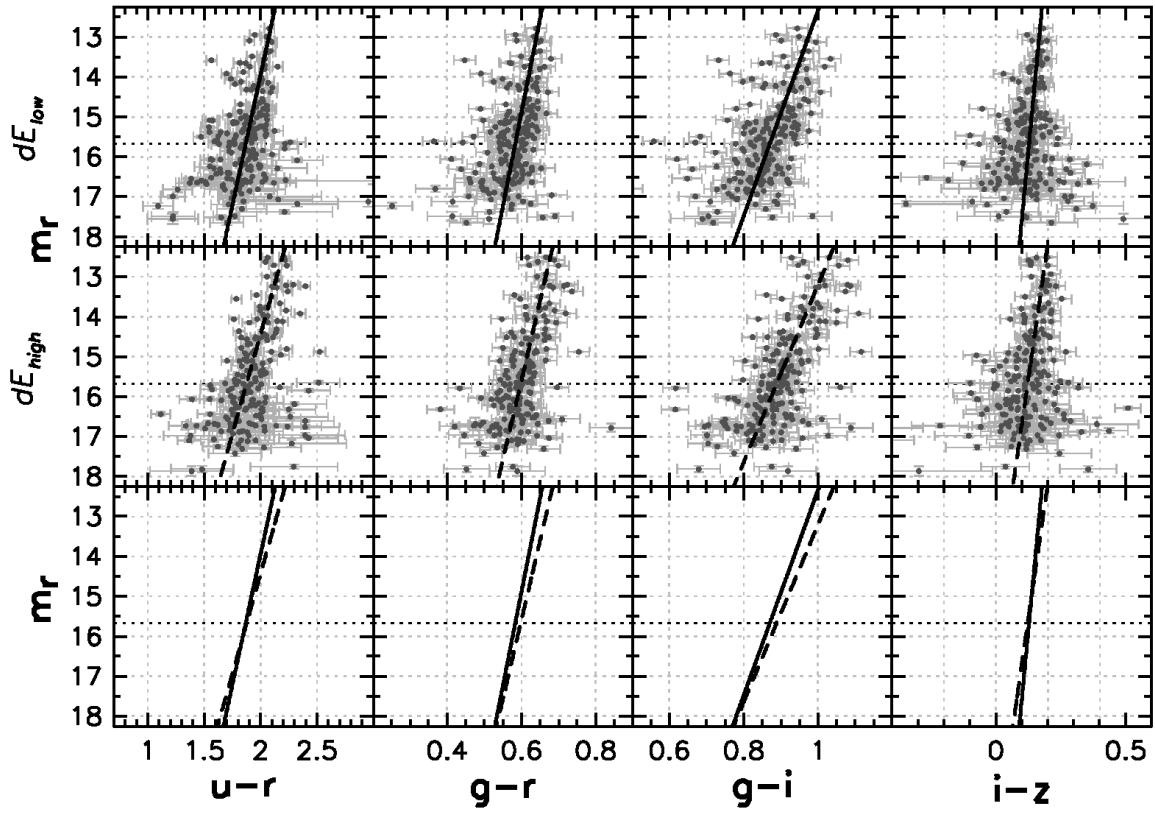} %{cmrdensall.eps}
  \caption{ Color-magnitude relations for different densities.
    Same as Figure~\ref{fig:cmrdens}, but for the full dE sample, excluding
  dE(bc)s.
  }
  \label{fig:cmrdensall}
\end{figure}

The resulting CMRs are shown
 in Figures~\ref{fig:cmrdens} and \ref{fig:cmrdensall}. The respective
 subsamples are denoted by a subscript 
 ``low'' for the low-density subsample, and ``high'' for the high-density
 subsample.  The pairwise
 statistical comparisons of the CMRs of low and high-density subsamples are
 given in Table~\ref{tab:tutdens}.

For all colors of $u-r$, $g-r$, and $g-i$, the dE(N)s, dE(nN)s, and the full dE
sample, the derived CMR at lower densities is consistently steeper than at higher
densities. Typically, the CMRs 
intersect at the fainter magnitudes of our sample, and with increasing
magnitude, galaxies are on average redder in the high-density
regime. In $i-z$, the low-density subsamples also follow a slightly
steeper CMR, but the intersection occurs at brighter magnitudes.
Note that, typically,
any correlation with \emph{projected} density would be even stronger with
true \emph{volume} density, since projection always causes some objects to
apparently lie close to the center that are actually situated in front of or
behind the center.

The difference between the
low and high-density subsamples of our full dE sample is significant in $u-r$
(probabilities for a common underlying distribution $\le$3.0\% for all
apertures, see Table~\ref{tab:tutdens}), and even
more significant in $g-r$ ($\le$0.9\% for all apertures) and $g-i$
($0.0\%$ for all apertures). 
The dE(N)s alone also display significant differences between the two density
regimes, namely for the intermediate apertures in $u-r$ (probability
$\le$2.0\%)  and $g-i$ ($\le$2.2\%). In $g-r$, the probability is still
lower than 6.5\%.
The fact that the
percentages are not as low as for the full dE sample, and that significant
differences are not found for all apertures, is at least partly due to the
lower sample 
size as compared to the full sample. Although a density difference is also
seen for the dE(nN)s in Figure~\ref{fig:cmrdens}, it is not statistically
significant:  the lowest probabilities
for a common distribution occur in $g-i$ for the small aperture ($\le$6.7\%)
and the intermediate aperture ($\le 10.2\%$). Here, probably the subsample sizes
are too small and the scatter of the CMRs too large for the rather small
differences to be significant.

The dE(N)s and dE(nN)s are populating different density regimes
(Paper III) and display significantly different CMRs (see
Section~\ref{sec:subclasses}). Therefore, one might conjecture that
the difference between the density-selected subsamples of the \emph{full} dE
sample stems mainly from the difference between the two subclasses. 
The fact that the dE(N)s alone also show
significant density differences would argue against this hypothesis, but
we nevertheless conduct a further test. We first
construct a ``combined sample'', which differs from
the full sample in that it contains no dE(di)s. Its median local projected
density is 1.295, and divides the sample into halves that, initially, do
\emph{not} contain an equal number of dE(N)s and dE(nN)s. For the
low (high) density half of the combined sample, we then exclude
randomly that many dE(nN)s (dE(N)s) that the number of galaxies belonging to
each subclass becomes equal. This leads to a combined sample containing 86
objects of each subclass in the low-density part, and 55  in the
high-density part, adding up to 282 objects in total. This 
approach guarantees that differences between the resulting low- and
high-density CMRs cannot be caused by one of the subclasses dominating
over the other in one density regime.

\begin{figure}
  \epsscale{1.0}
 \plotone{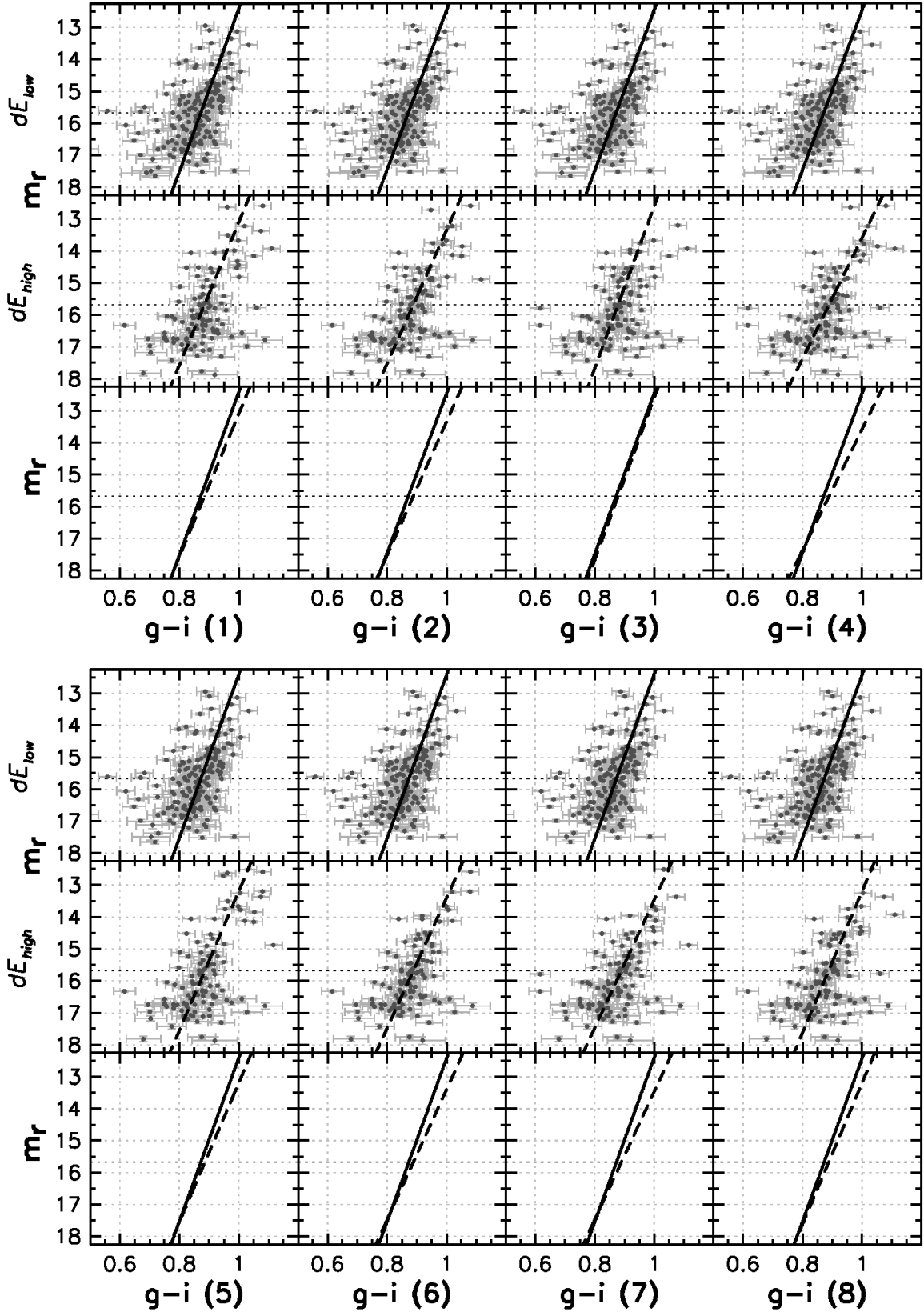} %{randomcmrs_my.eps}
  \caption{ Color-magnitude relations for different densities.
    Similar to Figure~\ref{fig:cmrdensall}, but showing the first 8 random
    realizations of the ``combined sample'' (see text), which contains equal numbers of
    dE(N)s and dE(nN)s in both its low- and high-density part. Colors were
    measured in $g-i$ using the half-light aperture.
  }
  \label{fig:cmrdensrandom}
\end{figure}

The
random exclusion of galaxies is repeated 100 times, and the
resulting CMRs are compared as previously described. We use the
intermediate aperture of $g-i$, where the density difference was found to
be significant for the dE(N)s alone (210 objects). The probability
for a common distribution has a median value of 1.2\% among our 100
realizations and exceeds 5\% in 11 cases and 10\% in 3 cases only. This
confirms that a density dependence of the CMR is present and is not faked
by an unequal combination of dE(N)s and dE(nN)s. Nevertheless, the
illustration of the first 8 realizations in Figure~\ref{fig:cmrdensrandom}
also leave room to speculate whether it is actually only the brighter dEs
whose color depends on density (and is reddest in high-density regions),
thereby affecting the fit to the CMR of all dEs.
A further subdivision of the dEs into bright and
faint subsamples, in addition to the subdivision by density, yields
significant differences for the bright part of the full dE sample in all
of $u-r$, $g-r$, and $g-i$ ($\le$2.4\% for all colors and
apertures). Still, at least for the small and intermediate apertures in
$g-i$, the statistical comparison also yields a significant difference for
the faint subsample ($\le$1.8\%). 
Among the dE(N)s and dE(nN)s, the lowest probability is the one for the
bright dE(N)s in $g-i$ for the small aperture, with $\le$6.6\%. 
An even larger sample size would be required to draw reliable conclusions
on the precise cause of the correlation between CMR and density.

So far, we did not consider the possibility that the density dependence
of the 
CMR could be caused by different average distances of the subsamples
along the line of sight. To test this, we consider the low and high-density
subsample of all dEs (excluding dE(bc)s) in $g-i$. At the
color value that the CMR of the high-density subsample has at the
bright reference magnitude, the CMR of the low-density subsample is
$0.62$ mag brighter. Assuming an average distance modulus of
$m-M=31.0$ mag, this would correspond to an average offset of the
subsamples along 
the line of sight of 4.5 Mpc. Moreover, both CMRs intersect at the
fainter magnitudes -- thus, the brighter dEs that are in regions of
higher projected density would have to be, \emph{on average}, 4.5 Mpc
closer to us than the fainter dEs in the same projected regions.

A difference in distance modulus of $1.0$ mag between the Virgo cluster
parts termed ``cluster A'' (i.e., basically
the part of the cluster that has a declination $\delta\gtrsim 10\arcdeg$, see
\citealt{virgokin}) and the western part of ``cluster B'' (southward of cluster A and
within a circular region around M\,49) was reported by \citet{gav99},
such that the latter part would be further away by 8.6 Mpc. Their study is
based on a combined sample of early-type and late-type galaxies. While the number
of dEs in this rather small region appears to be too low to
account for the differences that we see in the CMRs, we nevertheless test this
possibility by excluding all galaxies below a declination
$\delta<10\arcdeg$. However, the
difference between the low and high-density CMR is still equally large.
Within cluster A, \citet{gav99} give a dispersion of $0.45$ mag in their determination of
the distance modulus --- but they note that this value is
comparable to the nominal uncertainty in the methods used.
Based on the surface brightness fluctuations of 16 Virgo
dEs, \citet{jer04} found a distance dispersion of $\pm 1.45$ Mpc,
too low to be able to explain the observed differences. While
  their result is derived from a rather small sample, they noted that this
  dispersion, compared to the tangential extension of the cluster, implies a
  prolate spatial distribution of the galaxies that excellently agrees with
  other studies \citep{nei00,wes00}.
 We also point out
again that the color differences would require a \emph{systematic offset} of
4.5 Mpc, not just a distance dispersion. Furthermore, such an offset should
also be reflected in the magnitude distribution of the two subsamples --- yet
their median $r$ magnitudes differ by $0.07$ mag only, and their distributions
are very similar.
From
these considerations, it is clear that the significant correlation of
the CMR with local density is real and can not be explained with
differences in distance.

%%%%%%%%%%%%%%%%%%%%%%%%%%%%%%%%%%

\section{The scatter of the color-magnitude relation}
\label{sec:scatter}

If the observed scatter of the CMR would be solely due to measurement
uncertainties, no correlation should be present between the color residuals
in two different colors --- except, obviously, for colors that share a
certain band, like $u-r$ and $g-r$. If, on the other hand, a
significant true scatter was present (like, e.g., found by \citet{conII}
for early-type dwarfs in the Perseus cluster), one would expect at least some
correlation between the residuals: a dE that is intrinsically bluer in
$u-r$ than most of the other galaxies should typically not be
intrinsically redder in $g-i$ than most other dEs, since the wavelength
ranges covered by these colors 
overlap significantly. (It would, though, be ``allowed'' to be redder in
$i-z$.) Likewise, if the scatter was due to a spread in distance and
therefore in magnitude, most galaxies would fall on the same side of
the CMR in all colors.

We  therefore present in Figure~\ref{fig:cmroff} a pairwise comparison
of the color residuals for $u-r$ vs.\ $g-i$, $u-r$ vs.\ $i-z$, and $g-r$
vs.\ $i-z$. In each quadrant of each figure panel, we give the percentage of
objects that falls within it as black numbers.
A correlation is clearly seen for the dE(N)s and for the  full dE sample
in $u-r$ vs.\ $g-i$, for which more than twice as many objects lie
within the upper right and lower left quadrant than within the other two.
In $u-r$ vs.\ $i-z$, the correlation is
somewhat weaker, and again slightly weaker in $g-r$ vs.\ $i-z$. For the
dE(nN)s, only $u-r$ vs.\ $g-i$ shows a clear correlation. These results
indicate that there must be a significant true scatter of the CMR,
similar to the results of \citet{sec97} for the Coma
cluster.

Appendix~\ref{app:scatter} describes in more detail how we arrive
at the above conclusions. There, we also calculate that the
distance modulus dispersion of our galaxies would have to be  $\pm 0.61$
mag if it were the sole cause of the observed correlation. However, the
value measured by \citet{jer04} is only $\pm 0.177$ mag. Therefore, while
a certain distance scatter is naturally unavoidable, we conclude
that a significant \emph{intrinsic} color scatter must be present for our
  galaxies, with a strong correlation between the different colors.

A further question that can be addressed with our data is whether the
scatter of the CMR increases with decreasing luminosity. For example,
\citet{conII,conIII} found a significant increase of the CMR
scatter of early-type galaxies at magnitudes $\Mb\gtrsim -15$, roughly
corresponding to $\Mr\gtrsim -16$ and thus to $\mr\gtrsim 15$. In
order to quantify how the scatter changes with magnitude in our data, we
show in Figure~\ref{fig:scatterwithmag} the ratio of the rms of the color
residuals  about the CMR to the rms of the color errors, binned to
quartiles defined by the $r$ brightnesses of our full sample
(i.e., the four bins are separated at the previously defined bright
reference magnitude, ${m_\mathrm{r,bri.}}=14.77$, the median value,
$\mr=15.67$, and the faint reference magnitude,
${m_\mathrm{r,fai.}}=16.51$ mag). We use this ``rms-ratio'' in order to
detect any \emph{real} change in the scatter, i.e., one that is not caused
only by increasing measurement errors at fainter magnitudes. The values
shown were derived using the half-light aperture.

As can be seen from Figure~\ref{fig:scatterwithmag}, neither the dE(N)s, the
dE(nN)s, nor the full dE sample shows an increase in the CMR scatter at
fainter magnitudes in any color. In contrast, the scatter increases in
$u-r$ with \emph{increasing} brightness, and also does so in $g-i$ for the
\emph{brightest} bin of the dE(N)s. A trend for the latter is also seen
for the dE(nN)s in $g-i$: the second-brightest bin shows an
increase, but 
the brightest bin contains too few objects to be reliable.
These variations in the scatter can actually be seen qualitatively in
Figure~\ref{fig:cmr}, in particular the large scatter of the bright dE(N)s
in $g-i$. We can only speculate that this could be caused by a larger
spread in the stellar population properties of these objects, and that it
might be in some way related to the potential break in dwarf  galaxy
structure discussed  by \citet{bin91} at $M_{B_T}\simeq -16$. This magnitude
corresponds to $\mb=15.7$ given their $m-M=31.7$, and is thus roughly
equivalent to $\mr=14.7$. We can certainly rule out a significant increase
in the color scatter of the Virgo cluster dEs about the CMR down to $\mr
\approx 17$ mag (below which our sampling becomes more sparse), i.e.\ $\Mr
\approx -14$ mag, in contrast to what \citet{conIII} observed for  early-type
dwarfs in the Perseus cluster.

We further note that the overall scatter of the dE(nN)s is clearly larger
than that of the dE(N)s in all  colors except $i-z$
(Figure~\ref{fig:scatterwithmag}), again confirming the qualitative impression
from Figure~\ref{fig:cmr}.
%{\bf VON HIER}
 For this, there might be a simple explanation:
the color scatter of a sample of galaxies with a given spread in
stellar population age will decrease with time. This can be illustrated using
synthetic colors of a model stellar population that formed in a single exponentially
declining ($\tau = 1$ Gyr) peak of star formation, derived from population synthesis
models of \citet{bc03} with a metallicity of $\mathrm{[Fe/H]}=-0.64$ (see
Section~\ref{sec:stelpop}). We now consider the change in $g-i$ color of this
stellar population during 3-Gyr intervals from an age of 1.5 Gyr to 13.5 Gyr
(our adopted look-back time to the first star formation in the Universe, see
Section~\ref{sec:sub_modelsetup}). In the first interval, i.e., from 1.5 to
4.5 Gyr, the $g-i$ color reddens by 0.48 mag. In the next three intervals, it
reddens by only 0.18, 0.07, and 0.03 mag, respectively. Therefore, the color spread
of a sample of galaxies would, \emph{for a given age spread}, be smaller for
larger average ages. 
As we will show in the following section, the
dE(nN)s exhibit somewhat younger stellar populations than the dE(N)s, thus
providing a ``natural'' cause of the observed difference in CMR
scatter. Similarly, this could account for the rather large scatter of the
dE(bc)s in $g-i$ that is seen in Figure~\ref{fig:cmr}.
%{\bf BIS HIER}

%%%%%%%%%%%%%%%%%%%%%%%%%%%%%%%%%%
\section{Stellar populations}
\label{sec:stelpop}

\subsection{Population synthesis models}
\label{sec:sub_modelsetup}

We now attempt to use our color measurements for drawing conclusions about
the actual stellar population characteristics of our galaxies. For this
purpose, we construct several stellar population models, using the population
synthesis code from
\citet{bc03}. Following the recommendation of the authors, we use ``Padova
1994'' isochrones \citep{padova94}, as well as a Chabrier initial mass function
(IMF; \citealt{chabrier}).
We use
the \citeauthor{bc03} high resolution files, which rely on the STELIB
spectral library in the wavelength range $3200-9500$ \AA\ and on the
BaSeL 3.1 spectral library outside this range \citep[see][and
  references therein]{bc03}.
 All \citeauthor{bc03} models were calculated with fixed
metallicity, i.e., they do not take into account chemical enrichment.
 When adopting a concordance cosmology ($H_0=71$ km s$^{-1}$
Mpc$^{-1}$, $\Omega_{\rm m} = 0.27$, $\Omega_{\rm \Lambda} = 0.73$), it is now
13.67 Gyr since the Big Bang \citep{nedwright}. We thus assume that the
first stars have formed $\sim$13.5 Gyr ago \citep[also
  see][]{kas07}, which is  relevant 
for our model with constant star formation rate (see below). All of our models are
constructed for three of the seven available metallicities, namely
$Z=0.008$ ($\mathrm{[Fe/H]}=-0.33$), $Z=0.004$ ($\mathrm{[Fe/H]}=-0.64$),
and $Z=0.0004$ ($\mathrm{[Fe/H]}=-1.65$). 

Firstly, we construct a commonly used model, namely a stellar population
  formed through a single peak of star formation that exponentially decays
  with time (illustrated in the top panel of Figure~\ref{fig:sfr}). We
  choose decay times of $\tau = 1$ Gyr (the \expA\ model) and 
  2 Gyr (\expB\ model). The resulting model tracks are shown in the left column of
  Figure~\ref{fig:models} in various color-color diagrams. The tracks are
  curves of constant metallicity, and span a range of ages, from $1$ to $13.5$
  Gyr. Here, ``age'' means the time since the beginning of the star formation.

Secondly, we construct a model based on the study of \citet{dav88}, who
proposed a scenario for dE formation in which a dwarf irregular (dIrr)
experiences several short 
intense bursts of star formation, during which it would appear as blue compact
dwarf (BCD). These
bursts would increase the initial metallicity and
surface brightness of the dIrr, such that, after some time of passive
evolution and fading, it would eventually appear as dE. \citeauthor{dav88}
suggest ten bursts within a period of 1 Gyr, each one with a duration of 10 Myr.
We thus construct a corresponding model
(\burst\ model) with 10 bursts of star formation, the tenth one occuring 0.9
Gyr after the beginning of the first one (middle panel of
Figure~\ref{fig:sfr}). Each ``burst'' is a period of 
constant star formation rate (SFR) with a duration of 10 Myr. The corresponding
model tracks are shown in the middle column of Figure~\ref{fig:models}. Here,
``age'' means the time since the beginning of the first star formation
  burst.

Thirdly, we construct a model that is intended to represent the popular
scenario of dE formation through infall of a late-type galaxy into the
cluster \citep[e.g.][]{moo96}. The galaxy experiences a short period of enhanced star formation
through gas compression and similar effects, and is then stripped of its
remaining gas.
This model  (\cbten\ model) is defined by a constant
SFR since 13.5 Gyr ago, i.e., since the formation of the first
stars within our adopted cosmology (see above).
 Star formation is subsequently boosted by a factor of 100 during a short period of 10
 Myr, and is finally truncated (lower panel of Figure~\ref{fig:sfr}). The
corresponding model tracks, shown in 
the right column of Figure~\ref{fig:models}, are thus not tracks along different
formation ages, but along different truncation times. The analogue to a
\emph{young age} in the \expall\ and the \burst\ models therefore is a \emph{recent
  truncation} of star formation in the \cbten\ model, whereas a larger age
would correspond to a less recent truncation of star formation. Note that the
model tracks shown for the \expall\ and the \burst\ models actually consist of all
age points provided by \citet{bc03} (60 points or more), whereas the
truncation times of the \cbten\ model have only been calculated by us for the
points shown in the figure, which we then simply connected by lines.

The reason for plotting the model tracks in various combinations of 
color-color diagrams (Figure~\ref{fig:models}) is to find out which
combination of colors would come closest to breaking the famous
age-metallicity degeneracy (even if this cannot 
be achieved entirely with only optical photometry at hand). In addition, we
also need to take into account that the three bands $g$, $r$, and $i$ provide
the best S/N, whereas the S/N is a few times lower in $z$ and almost ten
times lower in $u$. Each color-color diagram basically relates a ``blue''
color (on the y-axis) to a ``red'' color (on the x-axis).

It can be clearly seen in Figure~\ref{fig:models} that the age-metallicity
degeneracy is very strong in the cases in which the $r$ 
band is used in the red color --- $i-z$ therefore is the obvious choice,
since it provides the reddest possible color. For the blue color, we first
note that $u-r$ does a better job in breaking the degeneracy than $u-g$, since
the color ranges covered 
by the models are larger for the former, but the errors are similar, if not
smaller, in $u-r$. However, given the considerably smaller errors, $g-r$ is
a useful alternative. We will therefore use both  the $u-r$ vs.\
$i-z$ and the $g-r$ vs.\ $i-z$ diagram in our analysis in
Section~\ref{sec:sub_modelapplication}. As an aside, note that for the \burst\ model,
the tracks are somewhat less squeezed together in \griz\ than in \uriz.

Before we proceed towards the application of our models, we briefly comment on
possible variants of our models. The main effect of weakening or amplifying
the final star formation burst in the \cbten\ model is a shift towards less
recent truncation times: in the case of no final burst and star formation
being truncated 0.5 Gyr ago, the colors are similar to the \cbten\ model and
truncation occuring 1 Gyr ago, or to a model with a three times stronger final
burst and truncation occuring $\sim$1.5 Gyr ago. Likewise, the dominating
effect of a truncation of star formation in the \expall\ models is a shift
towards younger ages.

%%%%%%%%%%%%%%%%%%%%%%%%%%%%%%%%%%
\subsection{Comparison of models and data}
\label{sec:sub_modelapplication}

It is generally
difficult to directly compare average color values of different dE
subsamples, since they are usually not sampled equally in luminosity, and
thus already have different colors due to the correlation of color and
magnitude. Therefore, we simply compare the color values of the linear fits
to the CMR of each
subsample, measured at the bright and faint ``reference magnitudes'', as
defined in Section~\ref{sec:subclasses}. This is shown in
Figures~\ref{fig:modelsplusdata} and \ref{fig:modelsplusdata_metal}, along
with the model tracks. In the left part 
of each figure, we use the \uriz\ diagrams for comparison with the models, while
the \griz\ diagrams are used in the right part. On each side, the respective
left column shows the color values at the bright reference magnitude, while
the right column shows the values at the faint reference magnitude. We
also compare the colors measured within our three different apertures: the
smallest symbol represents the small aperture, the largest symbol stands for
the large aperture, so that gradients can be recognized. Models are shown
as lines of constant metallicity in Figure~\ref{fig:modelsplusdata} and as
lines of constant age in Figure~\ref{fig:modelsplusdata_metal}.
 We note again that
the CMRs of the dE(bc)s and dE(di)s are only constrained by a handful of data
points at 
fainter magnitudes. The resulting color values should thus be taken with a
grain of salt, at least at the faint reference magnitude.

We first focus on the color values at the bright reference
magnitude. A systematic gradient can be seen for the dE(N)s, in the
sense that the stars in the inner regions of the galaxies have on average
higher metallicities and possibly also somewhat younger ages.
A very similar trend was also found for low-mass early-type dwarfs in the
Local Group \citep{har01}.
The dE(nN)s also show color differences with radius, but the
direction of the gradient is not as well-defined as for the dE(N)s: in the
 \uriz\ diagrams, the colors for the small and intermediate aperture would
 imply an almost equal metallicity but different age, whereas in the \griz\ diagrams they
 would suggest a roughly equal age. The colors within the large aperture
 correspond to a lower metallicity and older age as compared to the other
 apertures in both the \uriz\ and \griz\ diagrams.
For the dE(di)s, the colors within the large aperture also seem to hint at
a higher age, but no clear gradient can be seen between the apertures.
Again, this might be partly due to the only moderate sample
size. The dE(bc)s seem to have higher metallicities in the outskirts than in
the center --- here, near-infrared photometry would be desirable to guarantee
that the ``red color'' used in the diagrams is not affected by the light of
the young stars that are present in the center.

The figures illustrate that, within our simplified framework of models, the
stars of the dE(nN)s are, on average, either younger\footnote[6]{~Note
  that this statement does not constrain the actual 
  star formation history. If star formation began at the same epoch in both
  the dE(nN)s and dE(N)s, but lasted longer in the dE(nN)s, the stars of the
  latter would be younger on average. The same age difference could, however,
  be achieved if star formation began at a later epoch, but did not last
  longer than in the dE(N)s.}
 than those of the
dE(N)s, or of lower metallicity, or both. Unfortunately, the famous
age-metallicity degeneracy prevents us from making a more definite statement, but
in any case, the stellar population characteristics of dE(N)s and dE(nN)s
differ. 
 The dE(di)s appear to be on average slightly younger and/or slightly less
 metal-rich than the dE(N)s, a
result that is more pronounced in the \griz\ than in the \uriz\ diagrams. They
are, however, older and/or more metal-rich than the dE(nN)s. Overall, the metallicities of the
various subclasses lie at or below the track for $\mathrm{[Fe/H]}=-0.64$ --- however,
care must be taken with absolute numbers, given the assumptions and
simplifications on which the models are based.

At the faint reference magnitude, the differences between the subclasses
are smaller, and no systematic gradient within them can be seen. In the
\griz\ diagrams the
dE(nN)s still appear to be somewhat younger and/or less metal-rich than
the dE(N)s, but the difference is smaller than at the bright reference
magnitude, and it is even smaller in the \uriz\ diagrams. There is still a tendency for the
dE(nN)s to have lower metallicities and/or older ages in the outer regions
than in the center, while for the
dE(N)s, there is a slight reverse trend. However, we recall that the scatter in
$i-z$ at fainter magnitudes is rather large.

%%%%%%%%%%%%%%%%%%%%%%%%%%%%%%%%%%
\section{Summary and discussion}
\label{sec:discussion}

We have analyzed the colors of 413 Virgo cluster dEs by constructing
color-magnitude relations (CMRs) for different dE subclasses and
different local densities, as well as by comparing them to theoretical
colors from population synthesis models of \citet{bc03}.
 We found
significant differences between the CMRs of dE(N)s and dE(nN)s, as well
as between the CMRs at low and high local projected densities. The
models imply that the brighter dE(nN)s are younger than the dE(N)s and/or
have  lower metallicities. The dE(di)s are more similar to
the dE(N)s, yet still seem to be slightly younger and/or less metal-rich
on average. 
A significant intrinsic color scatter of the CMR is
present. The color residuals about the CMR are
correlated between different colors for the dE(N)s, and partly also
for the dE(nN)s, such that a galaxy falling on the blue side of the CMR
in one color also does so in the other color. We find no increase in the
color scatter at fainter magnitudes down to $\mr \approx 17$ mag ($\Mr
\approx -14$ mag).

The dE(N)s are consistent with having a nearly ``perfect'' intrinsic
correlation of colors, i.e., if the intrinsic $u-r$ color of a
dE(N) lies on the red side of the respective CMR, the same is true in almost
all cases for the intrinsic $i-z$ color. 
 This is particularly
interesting, since $u-r$ is more sensitive to the age of the stellar
population, while $i-z$ is sensitive to metallicity. A simple,
straightforward interpretation is that when the stars in a dE
are on average older than the typical value at that dE's luminosity,
then they are also more metal rich, and vice versa.
Assuming a direct correlation between luminosity and galaxy mass, we can
speculate that the intrinsic 
scatter of the CMR could, for a given initial mass,
reflect a spread in star formation rate or in the efficiency with
which gas is turned into stars, perhaps caused by environmental effects.
Neglecting other possible effects, a higher SFR at a given initial (gas) mass
would lead to stronger enrichment, i.e.\ to a higher
metallicity, than a lower SFR. The gas would be consumed more rapidly, thus
reaching the end of star formation earlier than with a lower SFR, and
consequently yielding older stars on average.

\subsection{Subclass colors and stellar populations}
\label{sec:sub_discuss1}

While we found in Section~\ref{sec:subclasses} that dE(nN)s and dE(N)s
follow different CMRs, we then discovered in
Section~\ref{sec:density} that the CMR depends on environmental
density. Since  dE(nN)s and dE(N)s populate different
density regimes (Paper III), we should compare the CMR of the dE(nN)s to
that of the \emph{low-density} subsample of dE(N)s: the median density
of the latter is 1.18 (in units of the logarithm of the number of galaxies per
square degree), and it is 1.20 for the dE(nN)s. In contrast, the
median density of the full sample of dE(N)s is 1.37. However,
the CMR of the low-density dE(N)s is still significantly different from
that of the dE(nN)s in $g-r$ and $g-i$; the probability for
a common distribution is $\le$0.1\% for the half-light
aperture. Likewise, the color difference between dE(N)s and dE(nN)s 
at the bright reference magnitude is $0.09$ mag in $u-r$ and $0.05$ mag in
both $g-r$ and $g-i$, using the half-light aperture. This difference only
changes by $0.01$ mag in $u-r$ and $g-i$ and even less in 
$g-r$ when considering only the low-density dE(N)s. Thus, the colors of the
 (bright) dE(nN)s do differ from those of the dE(N)s even if we allow
 for the different sampling in density.

The same test can be done for the dE(di)s: their densities (median
value 1.18) are also
much more comparable to those of the low-density dE(N)s than to those of the
full sample. While the color difference between the dE(di)s and the full dE(N)
sample at the bright reference magnitude is only $0.02$ mag in $u-r$, $0.03$
mag in $g-r$, and $0.04$ mag in $g-i$,
the statistical comparison of their CMRs yields significant differences
(Section~\ref{sec:subclasses}). This
changes when we compare the dE(di)s to only the low-density dE(N)
sample. Although the colors of the dE(di)s are still slightly bluer than
those of the low-density dE(N)s, none of these differences is
statistically significant.
 Whether or not this could indicate a close
relation between dE(di)s and dE(N)s despite their very different shapes
will be discussed in Section~\ref{sec:sub_discuss2}.

\citet{rak04} determined
ages and metallicities for 91 dEs in the Coma
and Fornax clusters, based on narrowband photometry. They derived ages
above 8 Gyr 
for the dE(N)s, which they found to be about 5 Gyr older than the
dE(nN)s. At least qualitatively, and in a relative sense, this would be
consistent with our results.
As for the metallicities, \citeauthor{rak04} found the dE(N)s to have
\emph{lower} metallicities than the dE(nN)s, conjecturing that
``globular clusters and dEN galaxies are primordial and have
metallicities set by external constraints such as the enrichment of
their formation clouds.'' This is not in agreement with our results for
the brighter magnitudes: there,
we find the metallicities of the dE(N)s to be either similar (in the inner
part of the galaxies, see Figure~\ref{fig:modelsplusdata}) or higher
(Figure~\ref{fig:modelsplusdata_metal}) than those of the dE(nN)s. Due to
the fact that the Virgo cluster is a dynamically
less relaxed structure than the Coma and Fornax clusters, it would be
particularly interesting to find systematic differences in the stellar
content of their galaxy populations.
%{\bf VON HIER}
However, it would be premature to conclude
that the dE(nN)s and dE(N)s in Virgo behave inversely to those in the other
clusters --- the sample of \citet{rak04} only comprises 10 dE(nN)s of the Coma
cluster and 9 dE(nN)s of the Fornax cluster, and moreover, the dE(nN)s show a
considerable color scatter in the metallicity-sensitive color $i-z$. Whether
or not these issues can account for the differences needs to be
analyzed in future work.
%{\bf BIS HIER}

The question of whether our fitted CMRs would be consistent
with being mainly a luminosity-metallicity relation can be qualitatively
addressed  by comparing the respective CMR values at the bright and faint
reference magnitude in Figure~\ref{fig:modelsplusdata_metal}. Overall
consistency with  this interpretation is present for all subclasses, but
we cannot exclude the additional presence of at least some age differences
with magnitude.
A general problem for the interpretation of the colors also is that the
color values for some data points lie at ``too large'' an age, i.e., they
fall above the age of the Universe for our model tracks. We thus need to emphasize again that many
simplifications entered the calculation of these tracks, like
the fact that the models are calculated at a fixed metallicity, or the rather
simple star formation histories that we consider. All our interpretations
of observed dE colors are always done within this simplified framework of
stellar population models.

\subsection{The minimum number of formation scenarios}
\label{sec:sub_discuss2}

On the basis of our observational analyses presented here and in the
previous papers of this series, we now attempt to
answer the question of how many different dE formation mechanisms there
must be \emph{at least} in order to explain the diversity of dE
subclasses. To start with, how confident can we be that the (flat)
dE(di)s do not belong to the same intrinsic (sub-)class as the (round) dE(N)s?
The distribution of intrinsic shapes of the brighter dE(N)s, as deduced in
Paper III, is rather broad and includes a significant number of
flat objects, even down to axial ratios of 0.3. Most of the range of
intrinsic axial ratios of the dE(di)s is thus covered by the range of values
of the dE(N)s. Moreover, 73\% of the dE(di)s are nucleated. While their
colors are somewhat different from the full sample of dE(N)s, the difference
is not anymore significant when compared only to the low-density dE(N)s,
which have the same median density as the dE(di)s
(Section~\ref{sec:sub_discuss1}).

If disk substructure, like spiral arms or bars, could only occur in the flattest dEs,
due to, e.g., the kinematical configuration of these objects, we would have
automatically selected only intrinsically flat galaxies in our search for disk
features (Paper I), and would obviously have found their
flattening distribution to be consistent with disk galaxies. The fact that
these show no central clustering could then be explained, for example, by the
much stronger tidal heating that a galaxy experiences in denser regions of the
cluster, leading to an earlier destruction of disk features \citep[cf.][]{mas05}. Similarly,
if dE(N)s and dE(di)s originated from a morphological transformation of
infalling late-type spirals through galaxy harassment \citep{moo96}, one could
imagine that the amount of transformation depended on how close the 
encounters with massive galaxies were that led to it --- and the probability
for close(r) encounters is obviously higher in the cluster center, leading to
rounder objects without disk features.

Why, then, are there almost no fainter dE(di)s? While
we concluded in Paper I that we most likely missed a
significant amount of dE(di)s at fainter magnitudes due to our
detection limits, we
also argued that the true number fraction \emph{does} decrease when
going to fainter objects. As a possible explanation, we can speculate
that disk substructure might be more likely to occur in more massive
galaxies, possibly connected to the presence of a certain amount of
rotational velocity.
These qualitative considerations demonstrate that dE(di)s and dE(N)s
could, in principle, have formed through the same formation process,
keeping our counter of necessary dE formation mechanisms at 1 for the
moment.

Could the dE(nN)s be also related to the dE(N)s and be formed by the same process?
 At least for the bright subsamples, 
the colors of dE(nN)s and dE(N)s differ significantly, indicating
younger average stellar ages of the dE(nN)s, or lower metallicities, or both (Section~\ref{sec:stelpop}).
This still holds true even when the different density distributions of
dE(nN)s and dE(N)s are accounted for (Section~\ref{sec:sub_discuss1}).
One might thus conjecture that the dE(nN)s simply formed more
 recently. The significantly flatter shapes of the bright dE(nN)s could
 then possibly be explained in the sense that they still need to
 experience several (further) encounters with massive galaxies, leading to
 further morphological transformation. One would, though, need to invoke
 another assumption, namely that the faint dE(nN)s, which are already
 significantly rounder than the bright dE(nN)s, were much more affected by
 the first tidal encounters, while the bright dE(nN)s were able to partly
 preserve their initial shape\footnote[7]{~However, note that the bright
 dE(nN)s might not even have experienced any tidal encounters yet, but could
 have been born as thick, puffy systems \citep{kau07}.}
 --- this could possibly be explained by the
 difference in mass. 

Another requirement of this scenario would be that the distribution of dE(nN)s
within the cluster, which is not centrally concentrated at all (Paper
III), would need to shift towards significantly larger local densities within
the next few Gigayears, in order to be similar to the distribution of today's
dE(N)s. However,
\citet{conI} derived a two-body relaxation time for the Virgo dEs of
much more than a Hubble time. Even violent relaxation, which
probably only applies for the case of infalling or
merging groups, would take at least a few cluster crossing times $t_{\rm cr}$,
with $t_{\rm cr}\approx 1.7$ Gyr for Virgo \citep{bos06}. It thus seems
difficult to reconcile these numbers with the  required dynamical process.

A further, obvious point is that nuclei would need to form soon in the
dE(nN)s. Perhaps nuclei are currently being formed in the centers of the
dE(bc)s where we are witnessing ongoing star formation (Paper II). Such a
scenario would lead to nuclei whose stars were clearly younger than the
vast majority of their host galaxies' stars. This would, however, be
hardly consistent with the results of \citet{acsvcs8}, who found "old to
intermediate-age populations" of dE nuclei, and of \citet{lotz04}, who
measured similar colors of nuclei and dE globular
clusters. An alternative would be that most nuclei form through
coalescence of globular clusters \citep[e.g.][]{oh00}. Yet in neither
case do we find an explanation for why the ratio of nucleated and
non-nucleated dEs should increase strongly with luminosity \citep{san85b}
if the latter were the immediate progenitors of the former.

Taken together, our observational results do not allow to explain dE
formation  with less than two different 
processes. We thus conclude by repeating the statement of \citet{vZe04a},
``we caution against single-channel evolutionary scenarios.'' Early-type
dwarfs are not a homogeneous class of objects, and we strongly recommend
to separately analyze the properties of dEs belonging to different
subclasses in any future study of dEs.

%________________________________________________________________

\acknowledgements
    We wish to express our gratitude to Mischa Vodi\v{c}ka for performing
    a preliminary photometric analysis of our dE sample, which helped us
    in optimizing our methods for image treatment and galaxy photometry.
    We thank Niranjan Sambhus for useful comments on statistical
    tests, and the referee for valuable suggestions. We gratefully
    acknowledge support by 
    the Swiss National Science Foundation through grants number
    200020-105260 and 200020-105535. 
    T.L.\ is supported within the framework of the Excellence Initiative
    by the German Research Foundation (DFG) through the Heidelberg
    Graduate School of Fundamental Physics (grant number GSC 129/1). T.L.\
    would like to thank Martin Altmann for useful SM macros.

    This study would not have been possible without the wealth of publicly
    available data from the SDSS.    
    Funding for the SDSS has been provided by the Alfred
    P. Sloan Foundation, the Participating Institutions, the National
    Science Foundation, the U.S. Department of Energy, the National
    Aeronautics and Space Administration, the Japanese Monbukagakusho,
    the Max Planck Society, and the Higher Education Funding Council
    for England. The SDSS Web Site is http://www.sdss.org/. 

    The SDSS is managed by the Astrophysical Research Consortium for
    the Participating Institutions. The Participating Institutions are
    the American Museum of Natural History, Astrophysical Institute
    Potsdam, University of Basel, Cambridge University, Case Western
    Reserve University, University of Chicago, Drexel University,
    Fermilab, the Institute for Advanced Study, the Japan Participation
    Group, Johns Hopkins University, the Joint Institute for Nuclear
    Astrophysics, the Kavli Institute for Particle Astrophysics and
    Cosmology, the Korean Scientist Group, the Chinese Academy of
    Sciences (LAMOST), Los Alamos National Laboratory, the
    Max-Planck-Institute for Astronomy (MPIA), the Max-Planck-Institute
    for Astrophysics (MPA), New Mexico State University, Ohio State
    University, University of Pittsburgh, University of Portsmouth,
    Princeton University, the United States Naval Observatory, and the
    University of Washington. 

    This research has made use of NASA's Astrophysics Data
    System Bibliographic Services, and of the NASA/IPAC Extragalactic
    Database (NED) which is operated by the Jet Propulsion Laboratory,
    California Institute of Technology, under contract with the
    National Aeronautics and Space Administration.

%________________________________________________________________

%\clearpage

%\appendix
\begin{appendix} 

   \section{The Petrosian uncertainty}
   \label{app:errors}

The uncertainty in the determination of the Petrosian SMA is
difficult to estimate, and even more so the subsequent uncertainty in the
total $r$ band flux and the corresponding half-light SMA. Apart
from proper masking of neighbouring or blended objects -- which we assume to
have been done sufficiently accurately -- the determination of the Petrosian SMA
is strongly affected by a possible over- or underestimation of the sky
level.
Should the sky
level be overestimated, as is the case in the measurements of the SDSS
photometric pipeline \citep[see][]{lis05}, the Petrosian SMA will be underestimated,
and vice versa. Thus, in the case of a sky level overestimation, the total
galaxy flux will be underestimated due to both the sky flux oversubtraction
\emph{and} the lower-than-actual Petrosian SMA within which the flux is measured. We
thus assume the ``Petrosian uncertainty'' to be of the same order as
the sky level uncertainty (see Section~\ref{sec:errors}), which we therefore
simply count twice in our calculation of total errors.

The Petrosian uncertainty is not directly relevant for the calculation of a
\emph{color} value of a given galaxy from the flux values of two bands,
since the aperture used is the same for both bands, independent of whether its
size was under- or overestimated. However, it is
relevant for comparing the color values of two different galaxies, since for
one of them, the half-light SMA might have been underestimated, but
overestimated for the other. In order to obtain a conservative error estimate
for the colors, we therefore decided to take into account the Petrosian
uncertainty in the same way as for the $r$ band total flux,
even if this might be somewhat too pessimistic.

  \section{Statistical comparison of two color-magnitude relations}
  \label{app:statistics}

Here we elaborate in more detail on our method of statistically comparing two
given color-magnitude relations
(Section~\ref{sec:sub_statcompare}). Obviously, one would consider two CMRs to 
be different if either 
their slopes, their zeropoints, or both were significantly different from
each other. Unfortunately, our method of linear fitting described in
Section~\ref{sec:sub_fitting} does not yield errors on these parameters --- but
even with a method that would yield formal errors, like least-squares fitting, it is
often the case that these errors are not realistic
\citep{nr}. The alternative to comparing the fitted lines would be to compare
the two underlying datasets directly. However, here we face the problem that
these might have different average magnitudes, like, e.g., the dE(N)s and the
dE(nN)s. For example, for two given datasets that follow exactly the same CMR
but probe different magnitude regimes, a two-dimensional \mbox{K-S} test would yield
a probability of zero that the datasets have the same underlying distribution,
simply because they are significantly different in the
magnitude-dimension. This is clearly not the sort of test we want to perform.

Given the above considerations, we decided on an approach that combines a
comparison of both the data points and the fitted lines of two datasets A and
B. First, we compute the color residuals of the data points A and the
data points B about the CMR A. This gives us two
distributions of a single parameter, namely the residual about the CMR A,
independent of magnitude. We can now compare these two distributions with each
other through a (one-dimensional) \mbox{K-S} test, which compares the cumulative
distributions, and through a Student's t-test for unequal variances, which
compares the means of the distributions. We then compute analogously the
residuals of the data points A and B about the CMR B. For the \mbox{K-S} test, we take
the average of both values of the \mbox{K-S} statistic $D$ \citep[which is the
maximum difference between two cumulative distributions, see][]{nr}, and
compute our final probability from it. For the t-test, we simply use the
larger of the two probabilities as our final probability.

The \mbox{K-S} test has the advantage that it is sensitive to
different distributions around the CMR, even if the mean of both was zero,
e.g.\ if both CMRs had different slopes but crossed each other at the middle
data point. The Student's t-test for unequal variances has the advantage of
taking into account the scatter of each dataset around the mean, which
implicitly includes the measurement errors.
Note that the
goodness of the linear fit is only taken into account implicitly to a small
extent: if one of the two linear relations was a rather bad fit or was based
on a small sample with large scatter, this would partially be counterbalanced
by the inverse comparison with the other CMR. If, however, both linear fits
were rather weakly defined, the resulting probabilities might not be too
useful.

\section{The scatter of the color-magnitude relation}
\label{app:scatter}

In order to obtain a rough estimate of the \emph{true} scatter of
the CMR (i.e., that is not caused by measurement errors), we can compare the
observed scatter with our measurement errors. The respective values are given
in Table~\ref{tab:cmr}: the column ``rms'' gives the root mean square of the
color residuals about the linear fit to the CMR, while the column ``E-rms''
gives the root mean square of the errors. If the observed scatter was solely
due to measurement errors, these two values should be similar, at least for
the larger (i.e., statistically robust) dE subsamples. The ratio of the two
values, which is given in the last column of the table, is thus an indirect
measure of the true scatter; we refer to it as the ``rms-ratio''.
The \emph{true} scatter can be a combination of an
\emph{intrinsic} color scatter and a distance spread; we shall consider the
latter at the end of this section.

We first note that for most CMRs, the rms-ratio increases with aperture
size. As for a possible intrinsic color scatter, it can hardly be deduced whether, and how
strongly, it increases from the inner to the outer parts of the dEs. As for our
measurement errors, we know that they consist of one part that remains constant
with aperture size, namely the error on the photometric calibration zeropoint
(``calibration uncertainty'', see Section~\ref{sec:errors}), and another part
that increases with aperture size, namely the S/N uncertainty and the sky
level uncertainty. It thus appears likely that an overestimation of the
calibration uncertainty causes, at least partly, the observed increase in the
rms-ratio with aperture size: due to this uncertainty, the measurement errors
do not approach zero with increasing S/N, but instead reach a finite
value. If this value -- which is provided directly by the SDSS for each color
-- would be overestimated, the rms-ratio would fall below a value of 1 for
high-S/N measurements, which is indeed the case for some of the small-aperture
CMRs, like e.g., for the bright dE(N)s in $i-z$. Again, we are not able to
tell how much of the  increase in the rms-ratio is an intrisic effect, but the
above considerations suggest that the photometric calibration might be at
least somewhat better than estimated.

We now attempt to derive an estimate for how large the true scatter is
with respect to our measurement uncertainties, i.e., which of the two
dominates the observed scatter. Given the above findings, we focus only on the
rms-ratios for the large-aperture CMRs, in order to not underestimate the
true scatter. When we consider only the larger dE subsamples, i.e., the
bright and faint dE(N)s, the faint dE(nN)s, and the full samples of dE(N)s,
dE(nN)s, and all dEs, we find that the rms-ratio always lies between 1.3 and
2. With the simplifying assumption that the rms of the true 
scatter and the rms of the measurement uncertainties add quadratically to
yield the observed rms, these values would imply that the rms of the true
scatter lies between 0.8 and 1.7 of that of the measurement
uncertainties.

\begin{figure}
 \epsscale{1.0}
 \plotone{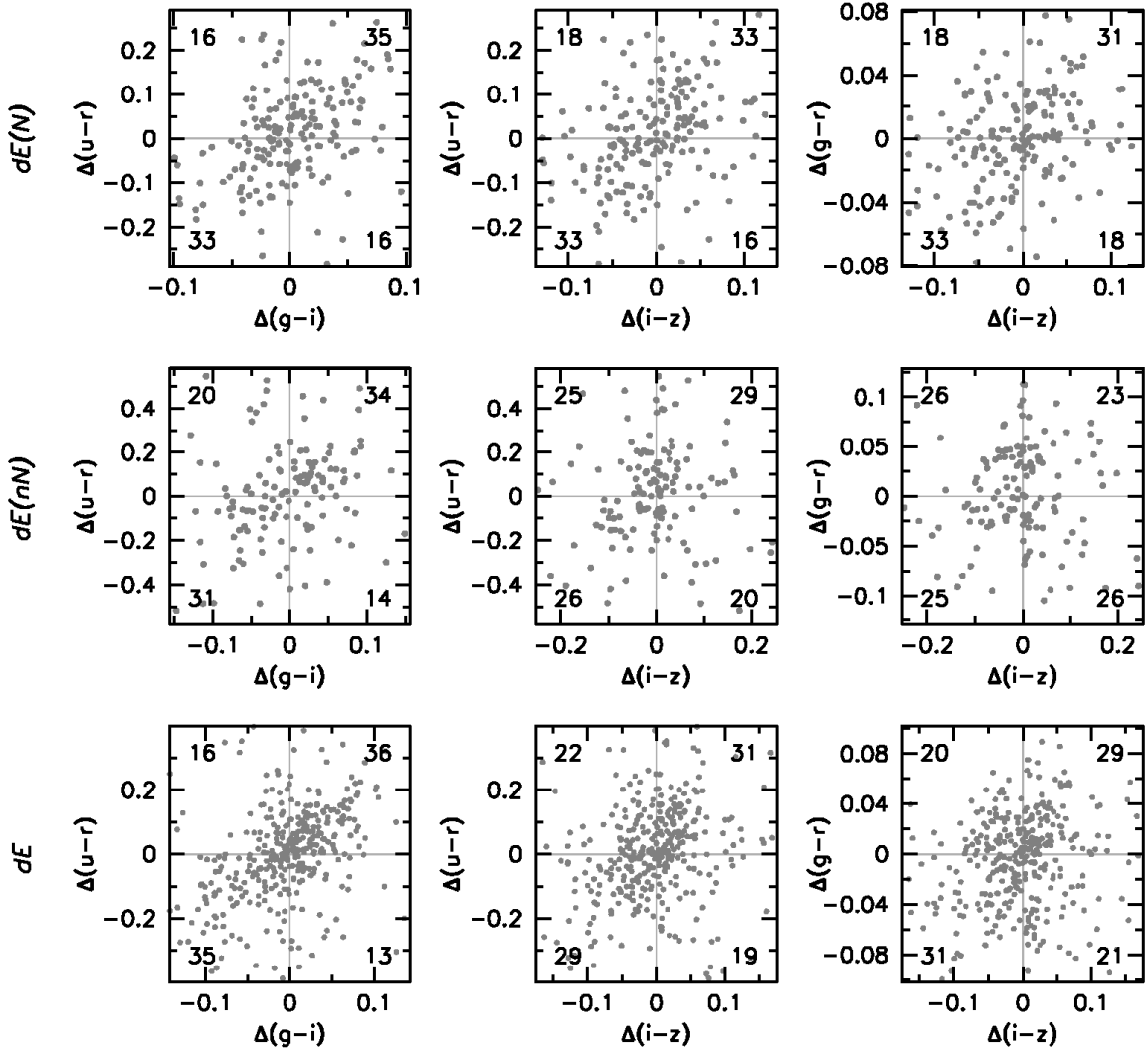} %{cmroff.eps}
 \caption{ Correlation of colors.
   Pairwise comparison of the color residuals about the respective CMRs, for
 colors measured within the half-light aperture. For a given galaxy,
 $\Delta(u-r)\,:=\,(u-r)\,-\,(u-r)_{\rm CMR}$, where $(u-r)$ is the galaxy's
 color, and $(u-r)_{\rm CMR}$ is the color value of the linear fit to the
 CMR at the $r$ magnitude of the galaxy; same for the other colors. The black
 numbers in the corners of the diagrams give the percentage of galaxies within
 the respective quadrant, bordered by the grey lines.
 }
 \label{fig:cmroff}
\end{figure}

\begin{figure}
 \epsscale{0.8}
 \plotone{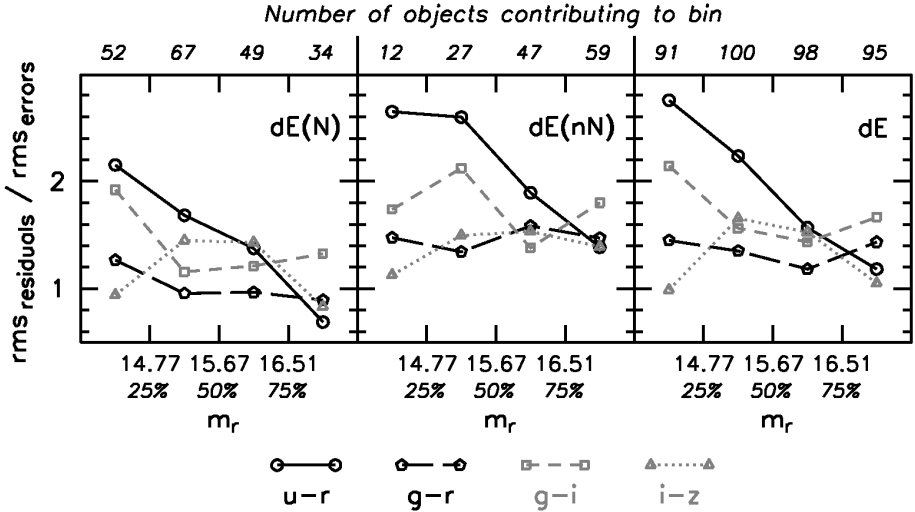} %{scatterwithmag.eps}
 \caption{ Relation of CMR scatter and magnitude for different
   colors. Shown is the ratio of 
   the rms of the color residuals  about the CMR to the rms of the color
   errors, binned to quartiles defined by the $r$ brightnesses of our full
   sample. Quartiles are separated at ${m_\mathrm{r,bri.}}=14.77$,
   $\mr=15.67$, and ${m_\mathrm{r,fai.}}=16.51$ mag (see text) as labelled
   below each panel; the x-axis
   thus does not follow a linear scale. Values were derived using the
   half-light aperture.
 }
 \label{fig:scatterwithmag}
\end{figure}

\begin{figure}
  \epsscale{0.5}
 \plotone{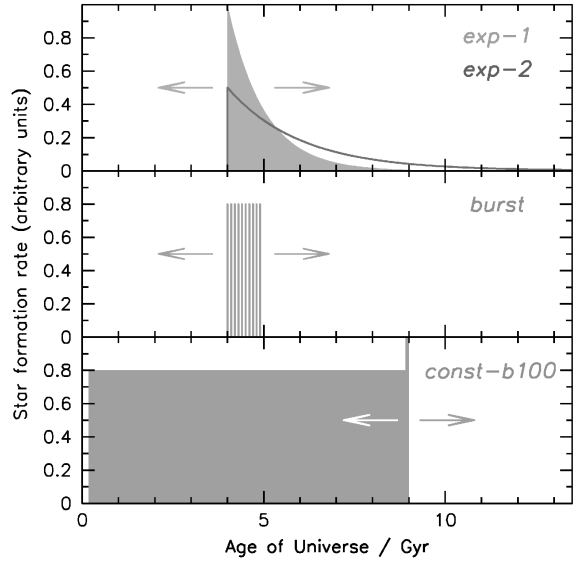} %{modelsketch_multi_my.eps}
  \caption{ Star formation histories.
    Illustration of the star formation histories used for our
    population synthesis models. \emph{Top panel:} The \expall\
    models. \emph{Middle panel:} The \burst\ model. \emph{Lower panel:}
    The \cbten\ model, in which the final burst reaches a SFR of 80 in the
    units adopted here. See text for details.
}
  \label{fig:sfr}
\end{figure}

\begin{figure}
  \epsscale{0.8}
 \plotone{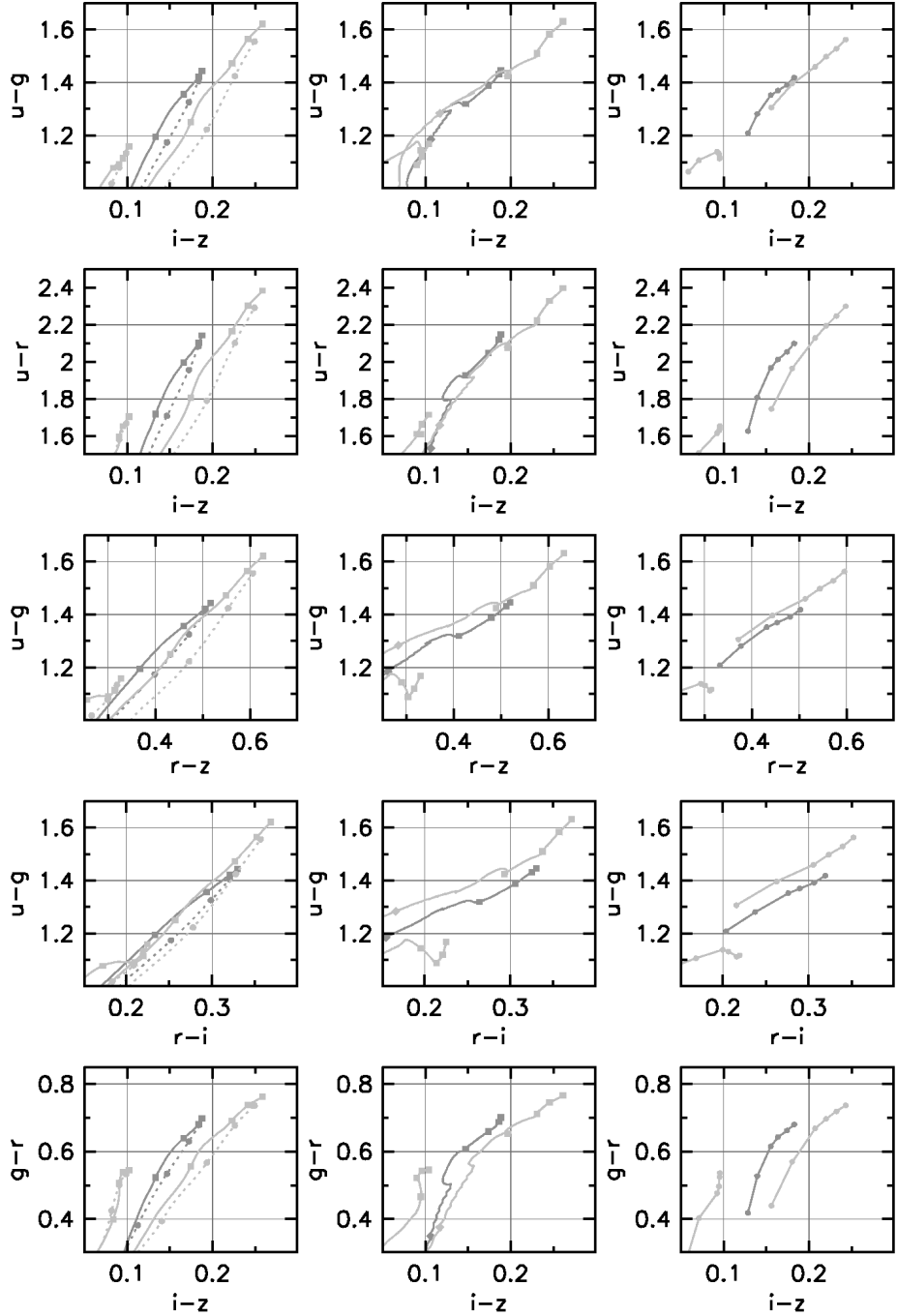} %{models2.eps}
  \caption{\small Population synthesis models.
  Shown are theoretical galaxy colors along \citeauthor{bc03} model tracks of
  constant metallicity, namely $Z=0.008$ ($\mathrm{[Fe/H]}=-0.33$, rightmost track,
  light grey), $Z=0.004$
  ($\mathrm{[Fe/H]}=-0.64$, middle track, dark grey), and $Z=0.0004$ ($\mathrm{[Fe/H]}=-1.65$,
  leftmost track, light grey).   
 The left column shows our \expall\ model, i.e., a star formation peak that exponentially declines
  with a decay time of $\tau = 1$ Gyr (solid lines) and 2 Gyr (dotted
  lines). The middle column shows our \burst\ model, i.e., 10 short bursts of
  star formation occuring within 1 Gyr. For both the \expall\ and the \burst\
  models, each model track reaches from an age of 1 Gyr (lower end, typically
  outside the plotting range) to 13.5 Gyr (upper end). Ages are marked at
  4.5, 7.5, 10.5, and 13.5 Gyr with filled squares or filled circles (\expB\
  model). For the \burst\ model, we give another age mark at 1.5 Gyr (filled
  diamond). The right column shows our \cbten\ model, i.e., constant star
  formation that is truncated at a certain epoch, including a final burst (see
  Figure~\ref{fig:sfr}). For this model, the model 
  tracks actually are simply lines that connect the data points for a
  truncation of star formation that, from bottom to top, occured 0.5, 1, 2, 3,
  5, and 7 Gyr ago.
}
  \label{fig:models}
\end{figure}

\begin{figure}
  \epsscale{1.0}
  \plotone{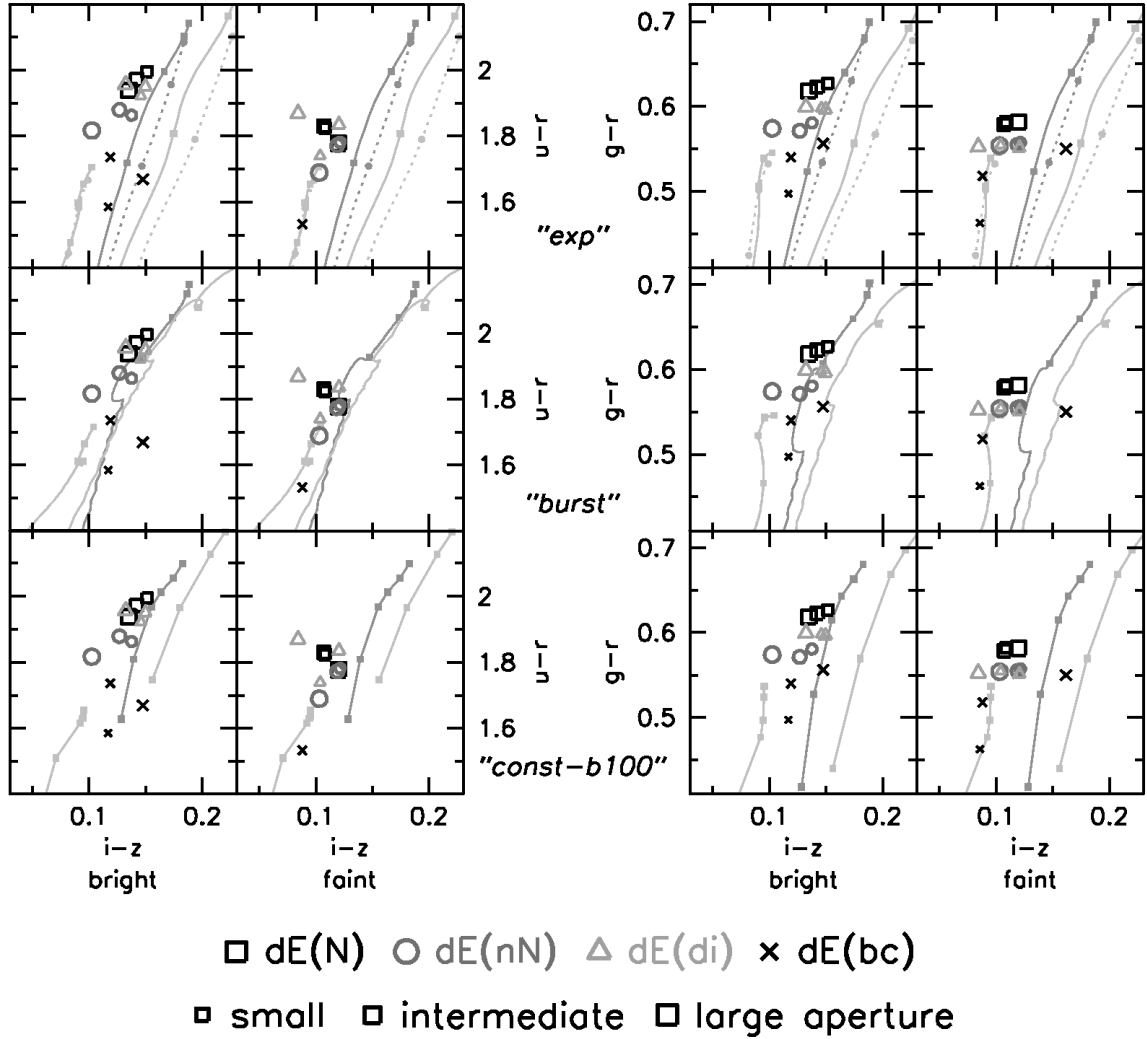} %{modelsplusdata.eps}
  \caption{Color and stellar population properties of the dE subclasses.
    Color values of the CMRs of dE(N)s (black squares), dE(nN)s (grey
    circles), dE(di)s (light grey triangles), and dE(bc)s (black crosses),
    measured 
    at the bright and faint reference magnitudes, as labelled below the
    diagrams. Small symbols indicate colors measured within the small
    aperture, intermediate-size symbols represent the intermediate aperture,
    and large symbols mark the colors within the large aperture, to
    illustrate systematic gradients. In the left part of the figure
    we show \uriz\ diagrams, while \griz\ diagrams are shown in the right
    part. Model tracks (as in Figure~\ref{fig:models}) are shown for the
    \expall\ models (top row), the \burst\ model (middle row), and the \cbten\
    model (bottom row). 
  }
  \label{fig:modelsplusdata}
\end{figure}

\begin{figure}
  \epsscale{1.0}
  \plotone{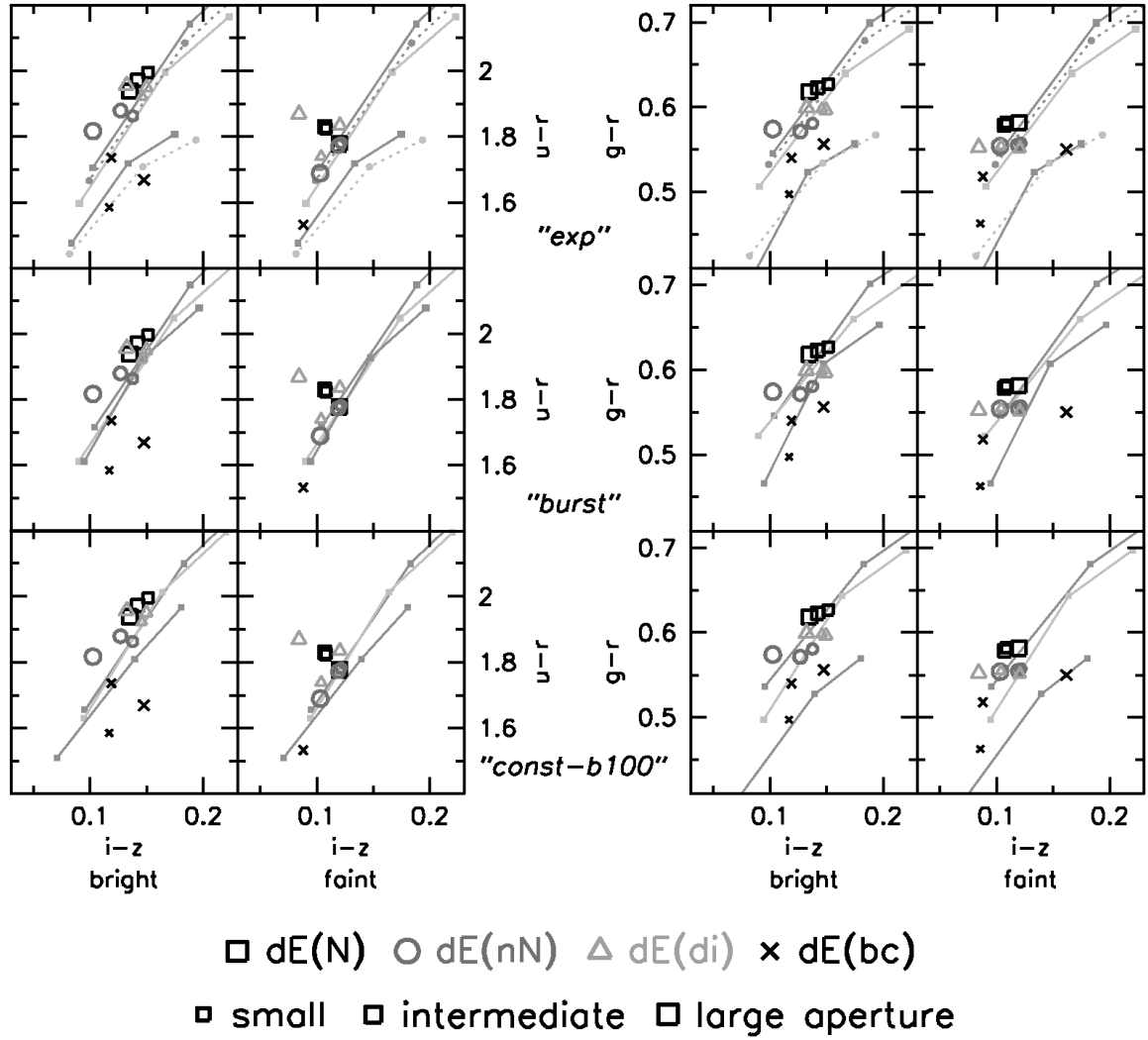} %{modelsplusdata_metal.eps}
  \caption{Color and stellar population properties of the dE subclasses.
    Same as Figure~\ref{fig:modelsplusdata}, but with model tracks now
    being lines of constant age (instead of constant metallicity). For the
    \expall\ and the \burst\ models, the tracks represent ages of 4.5,
    7.5, and 13.5 Gyr. For the \cbten\ model, the tracks connect points of
    constant truncation times, namely 1, 3, and 7 Gyr ago.
  }
  \label{fig:modelsplusdata_metal}
\end{figure}

We now attempt to estimate how the fraction of
galaxies within the different quadrants (Figure~\ref{fig:cmroff}) would be
distributed given a perfect 
correlation plus measurement errors. With ``perfect
correlation'', we mean that each galaxy falling on one side of the CMR in
one color falls on the same side of the CMR in the other color. We only care
about the \emph{direction} of the color offset, i.e., the sign of the
residual, not about its  absolute value.  In the following, we describe for the correlation of $u-r$
vs.\ $g-i$ how we simulate a CMR scatter. This simulation is done in the same
way for the other color pairs.

 In order to keep our approach simple, yet still instructive,
we refrain from choosing a certain analytic model distribution for the true
scatter of $u-r$ and $g-i$. Instead, we adopt the \emph{observed} distribution
of color values as one possible example for a true distribution.
 To guarantee that a given
galaxy falls on the same side of the CMR for both colors, simulating a
perfect correlation, we assign to the
 $g-i$ residual of each galaxy the sign of its corresponding $u-r$ residual (but keep its
absolute value). This represents our simulated distribution of
true\footnote[8]{~
 Our word choice here (``\emph{simulated} distribution of \emph{true}
 residuals'') might not be ideal, but we want to avoid speaking of
 an ``intrinsic distribution'', since the true scatter can be a combination of an
intrinsic color scatter and a distance spread. With ``true'', we only
mean that it is not caused by measurement uncertainties.}
 residuals. For simulating the distribution of measurement uncertainties, we use
again the observed distribution of color residuals, but redistribute these values
randomly among our galaxies, and assign them random signs.
 We then add these
values to the simulated true residuals, thereby yielding simulated
observed residuals. Note that in this case, the rms of the true scatter and
of the measurement uncertainties are obviously equal, since we used the same
absolute values. This can be altered by multiplying the simulated
true residuals with a certain factor.

The simulation was performed for the full dE sample (excluding
dE(bc)s), and was repeated 1000 times for all three independent color
pairs (i.e.,  $u-r$ vs.\ $g-i$, $u-r$ vs.\ $i-z$, and $g-r$ vs.\ $i-z$).
 We chose the rms value of the true scatter to be
0.8 and 1.7 times as large as the rms of the measurement uncertainties in
order to test our values deduced above. For simplicity, we term these two
cases the ``0.8-case'' and ``1.7-case'', respectively. For $u-r$ vs.\ $g-i$,
the resulting median fraction of galaxies within the upper right and lower
left quadrants is 63\% for the 0.8-case, and 72\% for the 1.7-case.
For both $u-r$ vs.\ $i-z$ and $g-r$ vs.\ $i-z$, the
corresponding values are 62\% and 72\%. Our observed value for $u-r$ vs.\
$g-i$ (70\%) lies between the two simulated cases and closer to the
1.7-case. For $u-r$ vs.\ $i-z$, the observed value is 60\%, and for $g-r$ vs.\
$i-z$, it is 59\% --- both values are only slightly below the simulated
0.8-case. These results confirm that our galaxy colors can well be explained
by a true scatter within the range deduced above, and that the
dEs, or at least the dE(N)s,
are consistent with exhibiting a strong correlation of color residuals
between  different colors.

We now need to investigate whether part of the
scatter of the CMR could be caused by a spread in distance, which would
lead to a certain scatter in magnitude, and would thus contribute to the
scatter of the CMR. 
We concentrate on the CMR of the dE(N)s
in $g-i$ for the large aperture. Here, the rms of our measurement errors is
$0.034$ mag (Table~\ref{tab:cmr}). Assuming the 0.8-case, the true rms scatter
woud be $0.027$ mag. The slope of the CMR is $-$0.044
(Table~\ref{tab:cmr}). Consequently, a scatter of $\pm 0.027$ mag around the
CMR would mean a scatter of $\pm 0.61$ mag in the distance
modulus, or
$^{+5.14}_{-3.88}$ Mpc
  at our adopted Virgo
  cluster distance of 15.85 Mpc ($m-M=31.0$ mag).
Such a huge (rms!) scatter is at variance with the observations ---  \citet{jer04} deduced a
  true dispersion in distance modulus of $\pm 0.177$ mag, corresponding to
  $^{+1.35}_{-1.24}$ Mpc (when using our distance modulus).
  When we adopt this observed value as rms distance scatter, the resulting
  color scatter for the CMR in question is $\pm 0.008$ mag.
 While this is rather small, it is nevertheless almost one third of the
  inferred true scatter in the 0.8-case (see above). However, the rms scatter from
  different uncertainties does most likely not add linearly --- if the
  ``distance uncertainty'' was added \emph{quadratically} to the intrinsic
  color scatter, it would have a very small effect only. In any case,
  while a small but finite distance scatter is naturally unavoidable,
  we conclude that a
  significant \emph{intrinsic} color scatter must be present for our
  galaxies, with a strong correlation between the different colors.

\end{appendix} 

%________________________________________________________________
 
%\clearpage

%\bibliography{virgodE_v3}
%\bibliographystyle{apj}

%________________________________________________________________

\clearpage
\begin{deluxetable}{llllllll}
  \tablecaption{
    Color-magnitude relation parameters.
    \label{tab:cmr}
  }
\tablewidth{0pt}
  \tablehead{
    \colhead{Type} & \colhead{Color}& \colhead{Aperture} & \colhead{Zeropoint} &
    \colhead{Slope} & \colhead{rms} & \colhead{E-rms} & \colhead{Ratio}\\
    &&&(mag)&&(mag)&(mag)&
  }
  \startdata
  dE(N) &   $u-r$ & small &   3.451 & $-$0.0986 &  0.146 &  0.139 &  1.046\\
  dE(N) &   $u-r$ & interm. &   3.168 & $-$0.0810 &  0.127 &  0.102 &  1.236\\
  dE(N) &   $u-r$ & large &   3.299 & $-$0.0922 &  0.177 &  0.113 &  1.567\\
  dE(N) &   $g-r$ & small &   1.008 & $-$0.0258 &  0.040 &  0.035 &  1.150\\
  dE(N) &   $g-r$ & interm. &   0.996 & $-$0.0253 &  0.033 &  0.032 &  1.017\\
  dE(N) &   $g-r$ & large &   0.931 & $-$0.0212 &  0.045 &  0.034 &  1.341\\
  dE(N) &   $g-i$ & small &   1.593 & $-$0.0450 &  0.050 &  0.035 &  1.429\\
  dE(N) &   $g-i$ & interm. &   1.549 & $-$0.0421 &  0.045 &  0.033 &  1.395\\
  dE(N) &   $g-i$ & large &   1.581 & $-$0.0441 &  0.062 &  0.034 &  1.797\\
  dE(N) &   $i-z$ & small &   0.516 & $-$0.0246 &  0.066 &  0.067 &  0.990\\
  dE(N) &   $i-z$ & interm. &   0.437 & $-$0.0200 &  0.066 &  0.056 &  1.168\\
  dE(N) &   $i-z$ & large &   0.262 & $-$0.0086 &  0.103 &  0.060 &  1.706\\
  \noalign{\smallskip}
  dE(nN) &   $u-r$ & small &   2.558 & $-$0.0470 &  0.225 &  0.224 &  1.004\\
  dE(nN) &   $u-r$ & interm. &   2.789 & $-$0.0616 &  0.255 &  0.165 &  1.548\\
  dE(nN) &   $u-r$ & large &   2.896 & $-$0.0731 &  0.317 &  0.193 &  1.641\\
  dE(nN) &   $g-r$ & small &   0.778 & $-$0.0134 &  0.055 &  0.044 &  1.253\\
  dE(nN) &   $g-r$ & interm. &   0.711 & $-$0.0095 &  0.056 &  0.038 &  1.488\\
  dE(nN) &   $g-r$ & large &   0.747 & $-$0.0117 &  0.068 &  0.042 &  1.642\\
  dE(nN) &   $g-i$ & small &   1.230 & $-$0.0246 &  0.070 &  0.044 &  1.565\\
  dE(nN) &   $g-i$ & interm. &   1.268 & $-$0.0265 &  0.067 &  0.038 &  1.741\\
  dE(nN) &   $g-i$ & large &   1.305 & $-$0.0285 &  0.074 &  0.042 &  1.764\\
  dE(nN) &   $i-z$ & small &   0.276 & $-$0.0093 &  0.104 &  0.096 &  1.080\\
  dE(nN) &   $i-z$ & interm. &   0.194 & $-$0.0046 &  0.106 &  0.075 &  1.424\\
  dE(nN) &   $i-z$ & large &   0.098 & \hphantom{$-$}0.0003 &  0.159 &  0.095 &  1.665\\
  \noalign{\smallskip}
  dE(di) &   $u-r$ & small &   3.469 & $-$0.1047 &  0.150 &  0.062 &  2.433\\
  dE(di) &   $u-r$ & interm. &   2.924 & $-$0.0660 &  0.119 &  0.056 &  2.138\\
  dE(di) &   $u-r$ & large &   2.695 & $-$0.0501 &  0.145 &  0.059 &  2.440\\
  dE(di) &   $g-r$ & small &   0.970 & $-$0.0252 &  0.035 &  0.030 &  1.185\\
  dE(di) &   $g-r$ & interm. &   0.981 & $-$0.0261 &  0.034 &  0.030 &  1.151\\
  dE(di) &   $g-r$ & large &   0.997 & $-$0.0270 &  0.030 &  0.030 &  1.013\\
  dE(di) &   $g-i$ & small &   1.609 & $-$0.0493 &  0.056 &  0.030 &  1.876\\
  dE(di) &   $g-i$ & interm. &   1.603 & $-$0.0481 &  0.052 &  0.030 &  1.770\\
  dE(di) &   $g-i$ & large &   1.727 & $-$0.0566 &  0.055 &  0.030 &  1.826\\
  dE(di) &   $i-z$ & small &   0.503 & $-$0.0241 &  0.024 &  0.040 &  0.585\\
  dE(di) &   $i-z$ & interm. &   0.397 & $-$0.0167 &  0.033 &  0.042 &  0.784\\
  dE(di) &   $i-z$ & large &   0.539 & $-$0.0275 &  0.059 &  0.044 &  1.358\\
  \noalign{\smallskip}
  dE(bc) &   $u-r$ & small &   4.911 & $-$0.2251 &  0.184 &  0.049 &  3.771\\
  dE(bc) &   $u-r$ & interm. &   3.463 & $-$0.1169 &  0.135 &  0.057 &  2.388\\
  dE(bc) &   $u-r$ & large &   4.445 & $-$0.1879 &  0.127 &  0.050 &  2.533\\
  dE(bc) &   $g-r$ & small &   0.789 & $-$0.0198 &  0.078 &  0.030 &  2.576\\
  dE(bc) &   $g-r$ & interm. &   0.727 & $-$0.0127 &  0.066 &  0.030 &  2.198\\
  dE(bc) &   $g-r$ & large &   0.607 & $-$0.0035 &  0.058 &  0.030 &  1.930\\
  dE(bc) &   $g-i$ & small &   1.175 & $-$0.0295 &  0.116 &  0.030 &  3.811\\
  dE(bc) &   $g-i$ & interm. &   1.230 & $-$0.0296 &  0.094 &  0.030 &  3.121\\
  dE(bc) &   $g-i$ & large &   1.270 & $-$0.0304 &  0.078 &  0.030 &  2.556\\
  dE(bc) &   $i-z$ & small &   0.382 & $-$0.0180 &  0.030 &  0.045 &  0.670\\
  dE(bc) &   $i-z$ & interm. &   0.380 & $-$0.0177 &  0.019 &  0.039 &  0.482\\
  dE(bc) &   $i-z$ & large &   0.028 &  \hphantom{$-$}0.0081 &  0.033 &  0.044 &  0.750\\
  \noalign{\smallskip}
  dE &   $u-r$ & small &   3.213 & $-$0.0848 &  0.179 &  0.172 &  1.042\\
  dE &   $u-r$ & interm. &   3.176 & $-$0.0830 &  0.186 &  0.128 &  1.459\\
  dE &   $u-r$ & large &   3.425 & $-$0.1023 &  0.238 &  0.145 &  1.646\\
  dE &   $g-r$ & small &   0.966 & $-$0.0240 &  0.049 &  0.038 &  1.283\\
  dE &   $g-r$ & interm. &   0.957 & $-$0.0233 &  0.047 &  0.034 &  1.362\\
  dE &   $g-r$ & large &   0.946 & $-$0.0228 &  0.053 &  0.036 &  1.461\\
  dE &   $g-i$ & small &   1.521 & $-$0.0413 &  0.062 &  0.039 &  1.611\\
  dE &   $g-i$ & interm. &   1.492 & $-$0.0391 &  0.058 &  0.035 &  1.684\\
  dE &   $g-i$ & large &   1.539 & $-$0.0420 &  0.069 &  0.037 &  1.858\\
  dE &   $i-z$ & small &   0.448 & $-$0.0204 &  0.072 &  0.076 &  0.946\\
  dE &   $i-z$ & interm. &   0.420 & $-$0.0189 &  0.078 &  0.061 &  1.275\\
  dE &   $i-z$ & large &   0.367 & $-$0.0161 &  0.119 &  0.074 &  1.597\\
  \enddata
  \tablecomments{
    ``rms'' is the root mean square of the color
    residuals about the linear fit, ``E-rms'' is the root mean square of the
    color errors, and ``Ratio'' is the ratio of rms and E-rms. Clipped
  data points are not taken into account in the calculation of the rms.
  }
\end{deluxetable}

\clearpage
\begin{deluxetable}{llllll}
  \tablecaption{ Statistical comparisons of color-magnitude relations of
  early-type dwarf subclasses.
    \label{tab:tut}
  }
\tablewidth{0pt}
  \tablehead{
    \colhead{Type 1} & \colhead{Type 2} & \colhead{Color}& \colhead{Aperture}
    & \colhead{\mbox{K-S} test} & \colhead{t-test}\\
    &&&&(\%)&(\%)
  }
  \startdata
  dE(N) & dE(nN)             & ~$u-r$ & small       &  \hphantom{1}0.00     &  \hphantom{1}8.55\\
  dE(N) & dE(nN)             & ~$u-r$ & interm.     &  \hphantom{1}0.00     &  \hphantom{1}6.02\\
  dE(N) & dE(nN)             & ~$u-r$ & large       &  \hphantom{1}0.02     &  \hphantom{1}9.10\\
  dE(N) & dE(nN)             & ~$g-r$ & small       &  \hphantom{1}0.00     &  \hphantom{1}0.00\\
  dE(N) & dE(nN)             & ~$g-r$ & interm.     &  \hphantom{1}0.00     &  \hphantom{1}0.00\\
  dE(N) & dE(nN)             & ~$g-r$ & large       &  \hphantom{1}0.00     &  \hphantom{1}0.04\\
  dE(N) & dE(nN)             & ~$g-i$ & small       &  \hphantom{1}0.00     &  \hphantom{1}0.01\\
  dE(N) & dE(nN)             & ~$g-i$ & interm.     &  \hphantom{1}0.00     &  \hphantom{1}0.00\\
  dE(N) & dE(nN)             & ~$g-i$ & large       &  \hphantom{1}0.00     &  \hphantom{1}0.53\\
  dE(N) & dE(nN)             & ~$i-z$ & small       &  15.29        &  68.98\\
  dE(N) & dE(nN)             & ~$i-z$ & interm.     &  16.48        &  68.92\\
  dE(N) & dE(nN)             & ~$i-z$ & large       &  15.89        &  29.76\\
  \noalign{\smallskip}                                                                    
  dE(N) & dE(di)             & ~$u-r$ & small       &  \hphantom{1}0.33     &  \hphantom{1}1.69\\
  dE(N) & dE(di)             & ~$u-r$ & interm.     &  14.86        &  27.66\\
  dE(N) & dE(di)             & ~$u-r$ & large       &  15.43        &  80.99\\
  dE(N) & dE(di)             & ~$g-r$ & small       &  \hphantom{1}1.17     &  \hphantom{1}2.28\\
  dE(N) & dE(di)             & ~$g-r$ & interm.     &  \hphantom{1}0.40     &  \hphantom{1}1.00\\
  dE(N) & dE(di)             & ~$g-r$ & large       &  \hphantom{1}3.50     &  11.25\\
  dE(N) & dE(di)             & ~$g-i$ & small       &  \hphantom{1}0.05     &  \hphantom{1}0.38\\
  dE(N) & dE(di)             & ~$g-i$ & interm.     &  \hphantom{1}0.03     &  \hphantom{1}0.92\\
  dE(N) & dE(di)             & ~$g-i$ & large       &  \hphantom{1}0.74     &  \hphantom{1}1.60\\
  dE(N) & dE(di)             & ~$i-z$ & small       &  \hphantom{1}3.89     &  59.39\\
  dE(N) & dE(di)             & ~$i-z$ & interm.     &  \hphantom{1}0.66     &  16.25\\
  dE(N) & dE(di)             & ~$i-z$ & large       &  \hphantom{1}1.07     &  48.83\\
  \noalign{\smallskip}                                                                    
  dE(nN) & dE(di)            & ~$u-r$ & small       &  \hphantom{1}1.99     &  25.19\\
  dE(nN) & dE(di)            & ~$u-r$ & interm.     &  \hphantom{1}2.80     &  37.77\\
  dE(nN) & dE(di)            & ~$u-r$ & large       &  \hphantom{1}0.07     &  50.74\\
  dE(nN) & dE(di)            & ~$g-r$ & small       &  \hphantom{1}8.60     &  37.56\\
  dE(nN) & dE(di)            & ~$g-r$ & interm.     &  \hphantom{1}2.90     &  67.77\\
  dE(nN) & dE(di)            & ~$g-r$ & large       &  \hphantom{1}4.59     &  87.08\\
  dE(nN) & dE(di)            & ~$g-i$ & small       &  11.43        &  39.27\\
  dE(nN) & dE(di)            & ~$g-i$ & interm.     &  11.52        &  53.16\\
  dE(nN) & dE(di)            & ~$g-i$ & large       &  \hphantom{1}3.04     &  10.28\\
  dE(nN) & dE(di)            & ~$i-z$ & small       &  \hphantom{1}1.16     &  40.18\\
  dE(nN) & dE(di)            & ~$i-z$ & interm.     &  \hphantom{1}0.02     &  \hphantom{1}8.62\\
  dE(nN) & dE(di)            & ~$i-z$ & large       &  \hphantom{1}0.11     &  41.91\\
  \enddata
  \tablecomments{
    The CMRs of a pair of dE subclasses (1st and 2nd column) are compared for
    a given color (3rd column) and aperture (4th column). Probabilities for a
    common underlying distribution are derived from a \mbox{K-S} test (5th column)
    and a Student's t-test for unequal variances (6th column); see text for
    details.
    The corresponding CMRs for the intermediate aperture are shown in
    Figure~\ref{fig:cmr}.
  }
\end{deluxetable}

\clearpage
\begin{deluxetable}{lllllllll}
\tablewidth{0pt}
\tabletypesize{\small}  
  \tablecaption{
    Color comparison of early-type dwarf subclasses.
    \label{tab:medbrifai}
  }
  \tablehead{
    Color & Aperture & dE(N) & dE(nN) & dE(di) & dE(bc) & dE & dE(N)$_{\rm bri.}$ & dE(nN)$_{\rm bri.}$\\
    &&(mag)&(mag)&(mag)&(mag)&(mag)&(mag)&(mag)
  }
  \startdata
  \sidehead{Color values at the bright reference magnitude}
  %\sidehead{Color values at the}
  %\sidehead{bright reference magnitude}
  ~$u-r$ & small & 1.995 & 1.863 & 1.922 & 1.585 & 1.960   & 1.999 & 1.864\\
  ~$u-r$ & interm. & 1.971 & 1.879 & 1.950 & 1.736 & 1.950 & 1.980 & 1.832\\
  ~$u-r$ & large & 1.938 & 1.817 & 1.956 & 1.669 & 1.914   & 1.937 & 1.802\\
  ~$g-r$ & small & 0.627 & 0.580 & 0.598 & 0.497 & 0.611   & 0.627 & 0.583\\
  ~$g-r$ & interm. & 0.622 & 0.571 & 0.596 & 0.540 & 0.613 & 0.623 & 0.582\\
  ~$g-r$ & large & 0.618 & 0.574 & 0.599 & 0.556 & 0.608   & 0.620 & 0.578\\
  ~$g-i$ & small & 0.929 & 0.867 & 0.881 & 0.739 & 0.911   & 0.925 & 0.865\\
  ~$g-i$ & interm. & 0.928 & 0.876 & 0.892 & 0.793 & 0.914 & 0.929 & 0.878\\
  ~$g-i$ & large & 0.930 & 0.884 & 0.891 & 0.820 & 0.918   & 0.928 & 0.883\\
  ~$i-z$ & small & 0.152 & 0.138 & 0.146 & 0.117 & 0.148   & 0.154 & 0.135\\
  ~$i-z$ & interm. & 0.142 & 0.127 & 0.150 & 0.119 & 0.141 & 0.145 & 0.128\\
  ~$i-z$ & large & 0.135 & 0.103 & 0.132 & 0.148 & 0.129   & 0.133 & 0.105\\
  \sidehead{Color values at the faint reference magnitude}
  %\sidehead{Color values at the}
  %\sidehead{faint reference magnitude}
  ~$u-r$ & small & 1.823 & 1.781 & 1.740 & 1.193 & 1.812   & ... & ...\\
  ~$u-r$ & interm. & 1.830 & 1.772 & 1.835 & 1.532 & 1.806 & ... & ...\\
  ~$u-r$ & large & 1.777 & 1.690 & 1.868 & 1.341 & 1.736   & ... & ...\\
  ~$g-r$ & small & 0.582 & 0.557 & 0.554 & 0.463 & 0.569   & ... & ...\\
  ~$g-r$ & interm. & 0.578 & 0.555 & 0.551 & 0.518 & 0.573 & ... & ...\\
  ~$g-r$ & large & 0.581 & 0.554 & 0.552 & 0.550 & 0.569   & ... & ...\\
  ~$g-i$ & small & 0.850 & 0.824 & 0.795 & 0.688 & 0.839   & ... & ...\\
  ~$g-i$ & interm. & 0.855 & 0.830 & 0.808 & 0.741 & 0.846 & ... & ...\\
  ~$g-i$ & large & 0.853 & 0.835 & 0.792 & 0.767 & 0.845   & ... & ...\\
  ~$i-z$ & small & 0.109 & 0.121 & 0.104 & 0.086 & 0.112   & ... & ...\\
  ~$i-z$ & interm. & 0.107 & 0.119 & 0.121 & 0.088 & 0.108 & ... & ...\\
  ~$i-z$ & large & 0.120 & 0.103 & 0.084 & 0.162 & 0.101   & ... & ...\\
  \enddata
  \tablecomments{
    Color values of the CMRs at the bright reference magnitude,
    ${m_\mathrm{r,bri.}}=14.77$ mag, and the faint reference magnitude,
    ${m_\mathrm{r,fai.}}=16.51$ mag (see text), for the three different
    apertures (2nd column). The two rightmost columns refer  to the
    respective bright subsample of dE(N)s and dE(nN)s.
  }
\end{deluxetable}

\clearpage
\begin{deluxetable}{lllll}
  \tablecaption{
    Statistical comparisons of color-magnitude relations for different densities.
    \label{tab:tutdens}
  }
  \tablehead{
    \colhead{Type} & \colhead{Color}& \colhead{Aperture} & \colhead{\mbox{K-S} test}
    & \colhead{t-test}\\
    &&&(\%)&(\%)
  }
  \startdata
  dE(N)   & ~$u-r$ & small   & \hphantom{1}7.85   & \hphantom{1}4.02\\
  dE(N)   & ~$u-r$ & interm. & \hphantom{1}1.95   & \hphantom{1}0.56\\
  dE(N)   & ~$u-r$ & large   & 32.11   & \hphantom{1}3.87\\
  dE(N)   & ~$g-r$ & small   & 93.16   & 24.89\\
  dE(N)   & ~$g-r$ & interm. & \hphantom{1}6.43   & \hphantom{1}3.23\\
  dE(N)   & ~$g-r$ & large   & 24.55   & \hphantom{1}3.14\\
  dE(N)   & ~$g-i$ & small   & \hphantom{1}8.22   & \hphantom{1}2.97\\
  dE(N)   & ~$g-i$ & interm. & \hphantom{1}1.84   & \hphantom{1}2.18\\
  dE(N)   & ~$g-i$ & large   & \hphantom{1}4.83   & \hphantom{1}1.09\\
  dE(N)   & ~$i-z$ & small   & \hphantom{1}5.58   & 44.52\\
  dE(N)   & ~$i-z$ & interm. & 65.68   & 86.18\\
  dE(N)   & ~$i-z$ & large   & 74.50   & 72.22\\
  \noalign{\smallskip}                                               
  dE(nN)   & ~$u-r$ & small   & 23.48  & 35.72\\
  dE(nN)   & ~$u-r$ & interm. & 15.43  & 51.11\\
  dE(nN)   & ~$u-r$ & large   & 31.48  & 29.56\\
  dE(nN)   & ~$g-r$ & small   & 47.96  & 19.16\\
  dE(nN)   & ~$g-r$ & interm. & 68.69  & 41.28\\
  dE(nN)   & ~$g-r$ & large   & 35.95  & 52.74\\
  dE(nN)   & ~$g-i$ & small   & \hphantom{1}6.70  & \hphantom{1}2.13\\
  dE(nN)   & ~$g-i$ & interm. & 10.24  & \hphantom{1}4.26\\
  dE(nN)   & ~$g-i$ & large   & 18.33  & \hphantom{1}6.35\\
  dE(nN)   & ~$i-z$ & small   & 26.07  & 63.19\\
  dE(nN)   & ~$i-z$ & interm. & 61.61  & 70.99\\
  dE(nN)   & ~$i-z$ & large   & 47.96  & 73.07\\
  \noalign{\smallskip}                                               
  dE   & ~$u-r$ & small   & \hphantom{1}0.46      & \hphantom{1}0.66\\
  dE   & ~$u-r$ & interm. & \hphantom{1}3.01      & \hphantom{1}1.65\\
  dE   & ~$u-r$ & large   & \hphantom{1}0.67      & \hphantom{1}0.93\\
  dE   & ~$g-r$ & small   & \hphantom{1}0.92      & \hphantom{1}0.01\\
  dE   & ~$g-r$ & interm. & \hphantom{1}0.10      & \hphantom{1}0.00\\
  dE   & ~$g-r$ & large   & \hphantom{1}0.23      & \hphantom{1}0.00\\
  dE   & ~$g-i$ & small   & \hphantom{1}0.01      & \hphantom{1}0.00\\
  dE   & ~$g-i$ & interm. & \hphantom{1}0.00      & \hphantom{1}0.00\\
  dE   & ~$g-i$ & large   & \hphantom{1}0.02      & \hphantom{1}0.00\\
  dE   & ~$i-z$ & small   & 13.24      & 21.72\\
  dE   & ~$i-z$ & interm. & 64.87      & 55.68\\
  dE   & ~$i-z$ & large   & 86.48      & 52.66\\
  \enddata
  \tablecomments{
    For a given dE subclass (1st column), color (2nd column),  and aperture
    (3rd column), we compare the CMRs of the low and high-density subsamples
    (see text). Probabilities for a common
    underlying distribution are derived from a \mbox{K-S} test (4th column) and a
    Student's t-test for unequal variances (5th column).
  }
\end{deluxetable}

%________________________________________________________________
 
\clearpage

\clearpage

\clearpage

\clearpage

\clearpage

\clearpage

\clearpage

\clearpage
 
\clearpage
 
\clearpage

\end{document}